\documentclass[twoside,journal]{IEEEtran}
\usepackage[T1]{fontenc}
\usepackage[utf8]{inputenc}
\usepackage{verbatim}
\usepackage{float}
\usepackage{amsmath}
\usepackage{amsthm}
\usepackage{amssymb}
\usepackage{mathdots}
\usepackage{graphicx}
\usepackage[unicode=true,
 bookmarks=false,
 breaklinks=false,pdfborder={0 0 1},colorlinks=false]
 {hyperref}
\hypersetup{
 hidelinks}

\makeatletter

\providecommand{\tabularnewline}{\\}
\floatstyle{ruled}
\newfloat{algorithm}{tbp}{loa}
\providecommand{\algorithmname}{Algorithm}
\floatname{algorithm}{\protect\algorithmname}

\let\oldforeign@language\foreign@language
\DeclareRobustCommand{\foreign@language}[1]{%
  \lowercase{\oldforeign@language{#1}}}
\theoremstyle{plain}
\newtheorem{thm}{\protect\theoremname}
\theoremstyle{definition}
\newtheorem{example}[thm]{\protect\examplename}

\usepackage{fancyhdr}
\usepackage{amsfonts}
\usepackage{fixmath}
\usepackage{nicefrac}
\usepackage{resizegather}

\DeclareMathOperator*{\argmin}{arg\,min}
\usepackage{mdframed}
\newmdenv[ 
topline=false,
leftline=false,
rightline=false,
bottomline=false,
]{txtbox}

\usepackage{ragged2e}
\usepackage{booktabs}
\usepackage{array}

\usepackage{url}
\usepackage{cite}
\usepackage{cleveref}
\crefformat{equation}{(#2#1#3)}
\crefrangeformat{equation}{(#3#1#4) --~(#5#2#6)}
\newcommand{\ie}{\textit{i.e.}}
\newcommand{\eg}{\textit{e.g.}}
\newcommand{\Tau}{\mathcal{T}}
\newcommand{\pib}{\boldsymbol\pi}
\newcommand{\hl}[1]{\textcolor{blue}{#1}} 

\newcolumntype{P}[1]{>{\centering\arraybackslash}p{#1}}
\newcolumntype{M}[1]{>{\centering\arraybackslash}m{#1}}
\newcolumntype{L}[1]{>{\raggedright\let\newline\\\arraybackslash\hspace{0pt}}m{#1}}
\newcolumntype{C}[1]{>{\centering\let\newline\\\arraybackslash\hspace{0pt}}m{#1}}
\newcolumntype{R}[1]{>{\raggedleft\let\newline\\\arraybackslash\hspace{0pt}}m{#1}}
\usepackage{multirow}

\usepackage{algorithm}
\usepackage{algpseudocode}

\usepackage{etoolbox}
\usepackage{tikz}
\usetikzlibrary{tikzmark}
\usetikzlibrary{calc}

\errorcontextlines\maxdimen

\newcommand{\ALGtikzmarkcolor}{black}
\newcommand{\ALGtikzmarkextraindent}{4pt}
\newcommand{\ALGtikzmarkverticaloffsetstart}{-.5ex}
\newcommand{\ALGtikzmarkverticaloffsetend}{-.5ex}
\makeatletter
\newcounter{ALG@tikzmark@tempcnta}

\newcommand\ALG@tikzmark@start{%
    \global\let\ALG@tikzmark@last\ALG@tikzmark@starttext%
    \expandafter\edef\csname ALG@tikzmark@\theALG@nested\endcsname{\theALG@tikzmark@tempcnta}%
    \tikzmark{ALG@tikzmark@start@\csname ALG@tikzmark@\theALG@nested\endcsname}%
    \addtocounter{ALG@tikzmark@tempcnta}{1}%
}

\def\ALG@tikzmark@starttext{start}
\newcommand\ALG@tikzmark@end{%
    \ifx\ALG@tikzmark@last\ALG@tikzmark@starttext
    \else
        \tikzmark{ALG@tikzmark@end@\csname ALG@tikzmark@\theALG@nested\endcsname}%
        \tikz[overlay,remember picture] \draw[\ALGtikzmarkcolor] let \p{S}=($(pic cs:ALG@tikzmark@start@\csname ALG@tikzmark@\theALG@nested\endcsname)+(\ALGtikzmarkextraindent,\ALGtikzmarkverticaloffsetstart)$), \p{E}=($(pic cs:ALG@tikzmark@end@\csname ALG@tikzmark@\theALG@nested\endcsname)+(\ALGtikzmarkextraindent,\ALGtikzmarkverticaloffsetend)$) in (\x{S},\y{S})--(\x{S},\y{E});%
    \fi
    \gdef\ALG@tikzmark@last{end}%
}

\apptocmd{\ALG@beginblock}{\ALG@tikzmark@start}{}{\errmessage{failed to patch}}
\pretocmd{\ALG@endblock}{\ALG@tikzmark@end}{}{\errmessage{failed to patch}}
\makeatother

\usepackage{resizegather}

\usepackage{xcolor}

\newtheorem{theorem}{Theorem}\newtheorem{proposition}[theorem]{Proposition}

\theoremstyle{definition}
\newtheorem{remark}{Remark}\newtheorem{definition}{Definition}

\newtheorem{example}{Example}

\makeatother

\providecommand{\examplename}{Example}
\providecommand{\theoremname}{Theorem}

\begin{document}

\vspace*{\fill}

\begin{minipage}{1.0\textwidth}
	$\copyright$ \the\year{} IEEE.  Personal use of this material is permitted.  Permission from IEEE must be obtained for all other uses, in any current or future media, including reprinting/republishing this material for advertising or promotional purposes, creating new collective works, for resale or redistribution to servers or lists, or reuse of any copyrighted component of this work in other works.
\end{minipage}

\vspace*{\fill}
\thispagestyle{empty}
\newpage

\title{Distributed Multi-Object Tracking Under Limited Field of View Sensors}
\author{Hoa Van Nguyen, Hamid Rezatofighi, Ba-Ngu Vo, and Damith C. Ranasinghe
\thanks{Acknowledgement: This work is supported by the Australian Research
Council under Linkage Project LP160101177 and Discovery Project DP160104662.} \thanks{Hoa Van Nguyen and Damith C. Ranasinghe are with
the School of Computer Science, The University of Adelaide, SA 5005,
Australia (e-mail: {hoavan.nguyen,damith.ranasinghe}@adelaide.edu.au).}\thanks{Hamid Rezatofighi is with the Department of Data Science \& AI,
Monash University, Clayton VIC 3800, Australia (e-mail: hamid.rezatofighi@monash.edu).}\thanks{Ba-Ngu Vo is with the Department of Electrical and Computer Engineering,
Curtin University, Bentley, WA 6102, Australia (e-mail: ba-ngu.vo@curtin.edu.au).} 
}
\maketitle
\setcounter{page}{1}
\begin{abstract}
We consider the challenging problem of tracking multiple objects using a distributed network of sensors. In the practical setting of nodes with limited field of views (FoVs), computing power and communication resources, we develop a \textit{novel} \textit{distributed multi-object tracking algorithm}. To accomplish this, we first formalise the concept of label consistency, determine a sufficient condition to achieve it and develop a novel \textit{label consensus approach} that reduces label inconsistency caused by objects' movements from one node's limited FoV to another. Second, we develop a distributed multi-object fusion algorithm that fuses local multi-object state estimates instead of local multi-object densities. This algorithm:~i)~requires significantly less processing time than multi-object density fusion methods; ii) achieves better tracking accuracy by considering Optimal Sub-Pattern Assignment (OSPA) tracking errors over several scans rather than a single scan; iii) is agnostic to local multi-object tracking techniques, and only requires each node to provide a set of estimated tracks. Thus, it is not necessary to assume that the nodes maintain multi-object densities, and hence the fusion outcomes do not modify local multi-object densities. Numerical experiments demonstrate our proposed solution's real-time computational efficiency and accuracy compared to state-of-the-art solutions in challenging scenarios. We also release source code at \textit{\url{https://github.com/AdelaideAuto-IDLab/Distributed-limitedFoV-MOT}} for our fusion method to foster developments in DMOT algorithms. 
\end{abstract}

\markboth{IEEE Transactions on Signal Processing}{Nguyen \MakeLowercase{\textit{et al.}}: Distributed tracking under
Limited FoVs}

\IEEEpeerreviewmaketitle{}

\section{Introduction}
\IEEEPARstart {T}{h}e aim of Multi-Object Tracking (MOT) is to estimate an unknown and time-varying number of object trajectories from noisy sensor measurements. MOT is an integral component in a multitude of applications in diverse domains, including surveillance~\cite{blackman1999design}, robotics~\cite{mullane2011a,hoa2019jofr,hoa2020iros}, computer vision~\cite{cox1996an}, traffic monitoring~\cite{munz2010generic,reuter2017fast}, cell biology~\cite{Hoseinnezhad2012visual,rezatofighi2015multi}, and space exploration~\cite{souza2016target}.  However, MOT is complicated by a time-varying number of objects, false alarms, misdetections, and data association uncertainty, in addition to the uncertainty resulting from process and measurement noise~\cite{mahler2007statistical}. There are three notable frameworks, amongst a range of algorithms, for tackling an MOT problem: multiple hypotheses tracking (MHT)~\cite{reid1979algorithm},  joint probabilistic data association (JPDA)~\cite{blackman1999design}, and random finite set (RFS) ~\cite{mahler2007statistical}.

Recently, driven by progress in wireless communication and sensing technologies, sensor networks comprising of interconnected \textit{nodes or agents} with sensing, communication and processing capabilities have attracted considerable research interest~\cite{battistelli2013consensus,battistelli2014kull,li2019computationally}. Importantly, for MOT, a network of multiple sensing nodes addresses the practical and critical problem of limited observability of sensing modalities at a single node, especially where objects are distributed across large spatial regions~\cite{souza2016target}, for example, tracking space debris using a network of low-earth-orbit cube satellites~\cite{Persico2019cubesat}. Sensor networks enable the inference of more accurate trajectories by \textit{fusing information} (\eg, estimates or densities) of multi-objects from observations at individual nodes (with limited observability) in a scalable (with respect to the number of nodes), flexible and reliable (\ie, resilient to failures) manner~\cite{luo2006dist}. However, leveraging these benefits requires a \textit{distributed mode} of operation for MOT where each node operates independently, without any knowledge of the network topology. Consequently, due to the significant benefits possible from using sensor networks for multi-object tracking applications, distributed MOT (DMOT) has attracted researchers' growing interest in recent years.

DMOT is a nontrivial problem and entails additional challenges. In particular, a suitable fusion algorithm is required to combine \textit{common information} associated with multi-objects to achieve improved tracking accuracy whilst merging \textit{complementary information} to overcome the limited FoV of a single sensor node. In principle, optimal fusion can be achieved by preserving marginal and joint multi-object distributions from all nodes~\cite{liggins1997dist}. However, maintaining these distributions requires sharing common information among all nodes~\cite{mahler2000optimal,battistelli2013consensus}, which limits the flexibility and scalability of the network. To maintain flexibility and scalability, robust (but sup-optimal) fusion solutions have been developed to address the double-counting of information when the common information is unknown~\cite{uney2013dist}.

Recent fusion techniques for DMOT were mostly developed from the random finite set framework because it facilitates principled generalisation of (single-object) distributed estimation to the multi-object case~\cite{mahler2000optimal}. This framework offers a convenient notion of multi-object (probability)  density that enabled the development of a suite of multi-object filters, \eg,~ the probability hypothesis density (PHD)~\cite{mahler2003phd}, cardinalised PHD (CPHD) ~\cite{mahler2007cphd},  multi-Bernoulli (MB) ~\cite{mahler2007statistical,vo2009cardinality,williams2015an}, and labelled multi-Bernoulli (LMB)~\cite{reuter2014lmb}, generalised labelled multi-Bernoulli (GLMB)~\cite{vo2013glmb,vo2014glmb}, and multi-scan GLMB~\cite{vo2019multiscan} filters. Many of these filters have been adapted for distributed multi-object estimation via the concept of Generalised Covariance Intersection (GCI\footnote{GCI is also known as Chernoff fusion~\cite{cover2012elements,chang2010analytical}, Exponential Mixture Density~\cite{julier2006using,clark2010robust,uney2013dist} or Kullback-Leibler Average~\cite{battistelli2013consensus,battistelli2014kull,fantacci2015consensus}})~\cite{mahler2000optimal,hurley2002information}, \eg,~ PHD-GCI~\cite{uney2013dist}, CPHD-GCI~\cite{battistelli2013consensus}, MB-GCI~\cite{wang2016distributed}, LMB-GCI~\cite{fantacci2018robust}, and its variations ~\cite{li2019computationally,li2017robust,yi2020comp}.

When the nodes do not \textit{share the same Field of View (FoV}), which is invariably the case with distributed sensor networks, GCI-based filters tend to perform poorly ~\cite{uney2019fusion}. A remedy for the PHD-GCI filter was proposed using cluster analysis (CA-PHD-GCI)~\cite{li2020Guchong}. However, this approach requires sharing FoV information among nodes and does not generate \textit{object labels (identities)}, an important function of an MOT algorithm~\cite{blackman1999design}. Labelled GCI-based filters generate object labels but tend to suffer from label inconsistency (\ie, individual objects are assigned different labels by different nodes). While there are efforts to reduce label inconsistency for sensors \textit{without FoV limitations}~\cite{li2017robust,li2019computationally,yi2020comp}, the \textit{more practical problems of reaching label consensus and reducing label inconsistency in distributed fusion with limited FoV sensors have not been addressed}. In addition to GCI, which is based on log-linear geometric averaging of probability densities, the linear arithmetic average has also been explored in the PHD ~\cite{li2017clustering,li2019local,LiTian2019partial,LiTian2020a}, CPHD~\cite{Gao2020multiobject}, and MB filter~\cite{LiTian2020on}. However, the arithmetic averaging consensus for DMOT has not yet been developed. 

While current fusion algorithms based on multi-object densities offer elegant conceptual solutions, these methods require intensive computing resources and high communication bandwidth. Due to the combinatorial nature of MOT, each multi-object distribution is characterised by a huge number of parameters. The computing resources needed to calculate these parameters at the nodes and the bandwidth needed for communicating them to neighbouring nodes can be prohibitive for real-time DMOT with increasing numbers of networked nodes and objects~\cite{gao2019event}. 
The high computational time and bandwidth requirements can be circumvented by considering the distributed fusion of local multi-object state estimates instead of multi-object densities. The track-to-track fusion and association algorithms investigated in ~\cite{chong1990distributed,chang1997on,chong2000architectures,mori2002track,mori2003track,kaplan2008assignment,mori2014performance,tian2015track} assume no false tracks nor missed objects, \ie, the number of local tracks from any two nodes are the same~\cite{mori2014performance}. This assumption is not practical in problems involving a time-varying number of objects and/or networks with limited FoV sensors. Further, most of the discussed distributed fusion approaches require  \textit{feedback}~\cite{mori2012comparison}, \ie, the fusion outcome is used to update the local multi-object densities.

In this work, we propose an efficient distributed fusion algorithm for DMOT that accounts for practical limitations on computing and communication resources as well as sensors with different and limited FoVs \textit{without feedback (\ie, the local multi-object densities are not modified)}. Our solution fuses local multi-object state estimates, and thus, is agnostic  to  local  multi-object  tracking   techniques. It employs a \textit{novel label consensus} algorithm that reduces inconsistency in the estimated labels caused by objects moving from one node's limited FoV to another. The optimal solution to this problem is NP-hard (for more than $2$ nodes). Considering the real-time requirements in practical applications, we propose a tractable sub-optimal fusion solution that reduces label inconsistency. This strategy incurs far less computation time and bandwidth than multi-object filtering density-based solutions. It also achieves better tracking accuracy by considering the tracking errors over several scans using the Optimal Sub-Pattern Assignment (OSPA) metric with OSPA track-to-track distance or OSPA\textsuperscript{(2)} metric~\cite{beard2017ospa2,beard2018performance,beard2020a}. We also formalise the label consistency concept and derive a sufficient condition to achieve label consistency by exploiting the metric properties of the OSPA\textsuperscript{(2)}. To validate our proposed method's effectiveness, we benchmark its accuracy and fusing time against state-of-the-art GCI-based solutions for MOT in a series of numerical experiments. 

The paper is organised as follows. We define the problem and provide background on metrics in Section~\ref{sec:background}. Section \ref{sec:info_fusion} presents our proposed fusion method. Section \ref{sec:experiments} details numerical experiments, results and comparisons with GCI-based methods. Section \ref{sec:conclusion} discusses concluding remarks.  

\section{Background}

\label{sec:background} This section provides the necessary background on distributed sensor networks and multi-object error metrics.

\begin{figure}[!tb]
\centering \includegraphics[width=0.5\textwidth]{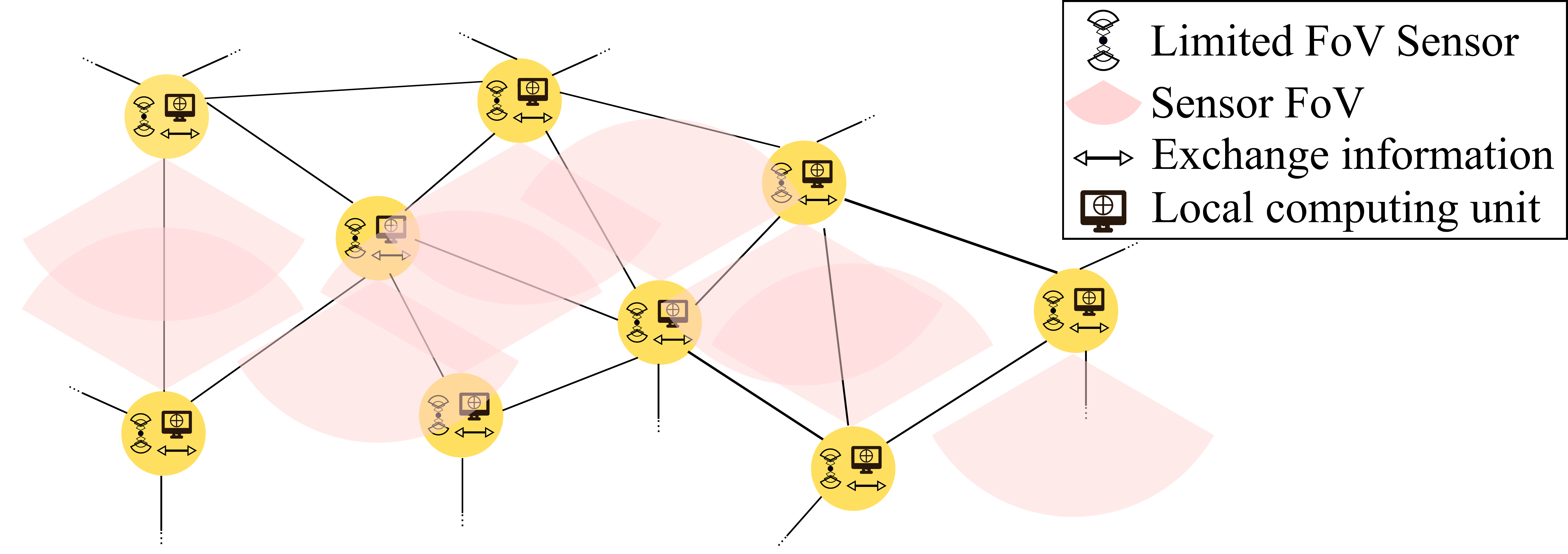}
\caption{A distributed sensor network system with limited FoVs.}
\label{fig:network_model} 
\end{figure}

\subsection{Distributed sensor network description} \label{sec:problem_description}
Fig.~\ref{fig:network_model} depicts a distributed heterogeneous network described by an undirected graph $G=(\mathcal{N},\mathcal{A})$, where $\mathcal{N}$ is the set of nodes and $\mathcal{A}\in\mathcal{N}\times\mathcal{N}$ is the set of arcs representing connections among nodes. An arc $(a,b)\in\mathcal{A}$ means that node $a$ can receive data from node $b$ and vice-versa. We denote by $\mathcal{N}^{(a)}\triangleq\{b\in\mathcal{N}:b\neq a\text{ and }(a,b)\in\mathcal{A}\}$ the set of neighbours of node $a$ from which data can be obtained. Each node is typically assigned with a unique ID. The set of all such IDs forms an ordered set (\eg,~ integers), and is assumed to be known by all the nodes.

Consider the task of monitoring a large area using the network of Fig.~\ref{fig:network_model} to detect and track an unknown and time-varying number of mobile objects. We assume that each node is equipped with a local computing unit capable of computing local multi-object state estimates, as well as a transceiver for communicating reliably with their neighbours through a limited range-and-bandwidth communication channel. In this context, each node can communicate its local multi-object state estimates to other nodes directly or indirectly  using a typical ad-hoc network or mesh network, for example, using an IEEE 802.16  standard. Each node is equipped with a limited field-of-view (FoV) sensor subjected to false alarms and misdetections. The network of interest has no central fusion node, and its nodes operate without knowledge of the network topology.

\begin{table}[!t]
\centering
\caption{List of variables/parameters}
\label{tab:parameters}
\begin{tabular}{|l|l|}
\hline
\textbf{Symbol} & \textbf{Variable/Parameter}      \\
\hline
$\mathcal{N}$               & set of nodes        \\
$\mathcal{A}$               & set of arcs among nodes \\
$\mathcal{N}^{(a)}$         & set of neighbours of nodes $a$ \\
$\mathbb{X}$                & single-object state space \\
$\mathbb{L}$                & label space \\
$\mathbb{T}$                & track space \\
$\mathbb{K}$                & discrete-time space \\
$\mathbb{I} = \mathbb{L} \times \mathcal{N}$                & global label space\\
$\mathbf{x} = (x,\ell)$               & labelled state \\
$\mathbf{X} \subset \mathbb{X} \times \mathbb{L}$               & labelled multi-object state \\
$\mathcal{L}(\mathbf{X}) \subset \mathbb{L}$               &  set of labels of $\mathbf{X}$ \\
$\mathbf{t} = (t,\ell)$               & labelled track \\
$\bar{\mathbf{t}} $               & labelled track estimate after fusion \\
$\mathcal{D}^{(\mathbf{t})} \subseteq \mathbb{K}$               & set of time instances that $\mathbf{t}$ exists \\
$\mathbf{T} \subset \mathbb{T} \times \mathbb{L}$               & set of labelled tracks \\
$\mathcal{T}(\mathbf{T}) \subset \mathbb{T}$               &  set of unlabelled tracks of $\mathbf{T}$ \\
$\mathbf{X}^{(a)} \subset \mathbb{X} \times \mathbb{I}$               & set of globally labelled state estimates of node $a$ \\
$\mathbf{L}^{(a)}  \subset \mathbb{I}$               & set of global labels of node $a$ \\
$\mathbf{T}^{(a)} \subset \mathbb{T} \times \mathbb{I}$               & set of globally labelled tracks of node $a$ \\
$\mathbf{X}^{(con)} \subset \mathbb{X} \times \mathbb{I}$               & set of consensed globally labelled state estimates \\ \hline
\end{tabular}%
\end{table}

At time $k$, an existing object is represented by a \textit{labelled state} $\mathbf{x}_{k}=\left(x_{k},\ell_{k}\right)$, where $x_{k}\in\mathbb{X}$ is its \textit{state} vector, and $\ell_{k}\in\mathbb{L}$ is a unique label consisting of the time of birth and an index to distinguish objects born at the same time. The set $\mathbf{X}_{k}\subset\mathbb{X\times L}$ of labelled states of existing objects (with distinct labels) is called the \textit{labelled} \textit{multi-object state} (the set $X_{k}\subset\mathbb{X}$ of states of existing objects is called the \textit{multi-object state}). We use the notation $|X|$ for the cardinality of a set $X$, and $\mathcal{L}(\mathbf{X})=\{\mathcal{L}(\mathbf{x}):\mathbf{x}\in\mathbf{X}\}$ for the set of labels of a multi-object state $\mathbf{X}$, where $\mathcal{L}:\mathbb{X}\times\mathbb{L}\rightarrow\mathbb{L}$ is the label projection defined by $\mathcal{L}((x,\ell))=\ell$. 

\begin{definition}
Given a (discrete) time window $\mathbb{K}\triangleq\{1,\dots,K\}$ from start time $1$ to end time $K$ of the scenario, the  \textit{track space} $\mathbb{T}$ is the space of all functions $t:\mathbb{K} \rightarrow \mathbb{X}$. Any element $t\in\mathbb{T}$ is called a \textit{track}, and the ordered pair $\mathbf{t}=(t,\ell)\in\mathbb{T\times L}$ is called a \textit{labelled track}.
\end{definition}

The set $\mathcal{D}^{(\mathbf{t})}\subseteq\mathbb{K}$ of time instances that the track $\mathbf{t}=(t,\ell)$ exists is the domain $\mathcal{D}^{(t)}$ of $t$, and the labelled state vector at time $k\in\mathcal{D}^{(\mathbf{t})}$ is given by $\mathbf{x}_{k}=(t(k),\ell)$. We also use the same notation for the label projection $\mathcal{L}((t,\ell))=\ell$, and  $\mathcal{L}(\mathbf{T})=\{\mathcal{L}(\mathbf{\mathbf{t}}):\mathbf{\mathbf{t}}\in\mathbf{\mathbf{T}}\}$, for $\mathbf{T}\subset\mathbb{T\times L}$. Further, let $\mathcal{T}:\mathbb{T}\times\mathbb{L}\rightarrow\mathbb{T}$ be the track projection defined by $\mathcal{T}((t,\ell))=t$ and $\mathcal{T}(\mathbf{T})=\{\mathcal{T}(\mathbf{\mathbf{t}}):\mathbf{\mathbf{t}}\in\mathbf{\mathbf{T}}\}$. 

\begin{figure}[!tb]
\centering \includegraphics[width=0.4\textwidth]{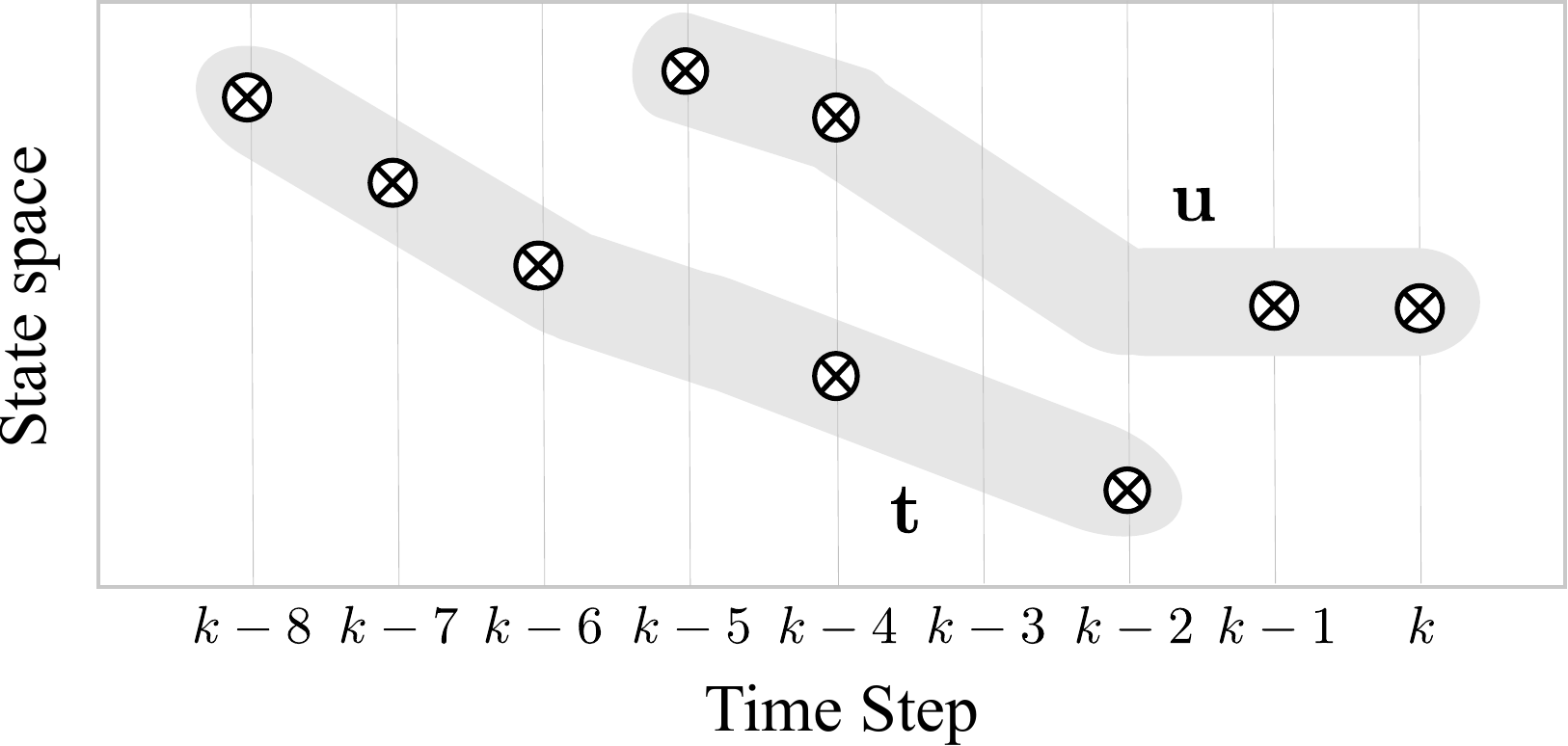}
\vspace{-0.2cm}
 \caption{A set of two fragmented tracks $\mathbf{t}$ and $\mathbf{u}$. }
\label{fig:track_def} 
\end{figure}

Two remarks are in order. First, the definition of labelled tracks meets the MOT requirement that tracks have \textit{labels (or identities)}~\cite{blackman1999design}. In practice, labels are important for identifying/referencing tracks; for example, given multiple tracks on display, it would be impractical for a human user to communicate with other users about a track by specifying every estimated position of its trajectory.  Second, the set $\mathcal{D}^{(\mathbf{t})}$ of instances that track $\mathbf{t}$ exists need not be consecutive (see Fig.~\ref{fig:track_def}), and hence encompasses the so-called \textit{fragmented tracks} generated by many MOT algorithms. A \textit{fragmented track} arises when the MOT algorithm declares it non-existent (possibly due to misdetections) in between instances where it (is declared to) exist, as illustrated in Fig.~\ref{fig:track_def}. When $\mathcal{D}^{(\mathbf{t})}$ consists of consecutive instances, track $\mathbf{t}$ is called a \textit{trajectory}.

Given a finite set $\mathbf{T}\subset\mathbb{T\times L}$ of labelled tracks with distinct labels, the labelled multi-object state at time $k\in\mathbb{K}$ is given by $\mathbf{X}_{k}=\cup_{(t,\ell)\in\mathbf{T}}\{(t(k),\ell)\}$. Note that the history $\mathbf{X}_{1:K}\triangleq\mathbf{X}_{1},...,\mathbf{X}_{K}$ of labelled multi-object states completely determines the set $\mathbf{T}$, and hence we use the notation $\mathbf{X}_{1:K}$ to denote the set of labelled tracks on the interval $\mathbb{K}$. We denote the restriction of the track $t$, labelled track $\mathbf{t}$, and set of labelled tracks $\mathbf{X}_{1:K}$ (or $\mathbf{T}$) on the window $\{j:k\}\subset\mathbb{K}$, by $t_{j:k}$ , $\mathbf{t}_{j:k}$, and $\mathbf{X}_{j:k}$ (or $\mathbf{T}_{j:k}$). Hereon, the discussions in this paper only concern the time window $\{j:k\}$, hence, for notational compactness, we drop the subscript $j:k$ when no confusion arises, \eg,~ $\mathbf{T}\triangleq\mathbf{T}_{j:k}$.

Instead of seeking consensus amongst local multi-object densities as pursued in many of the latest works~\cite{li2019computationally,li2020Guchong}, we are interested in reaching consensus amongst local multi-object state estimates from the set $\mathcal{N}$ of nodes.  At time $k$, each local node $a$ communicates a message which is an ordered pair $(a,\mathbf{X}_{k})$ comprising of the node's identity $a$ and the local labelled multi-object state estimate $\mathbf{X}_{k}$. Let $\mathbb{I}=\mathbb{L}\times\mathcal{N}$ be the global label space of the network, wherein each global label is unique across the network. Let $\mathbf{X}_{k}^{(a)}=\{(x,(\ell,a)):(x,\ell)\in\mathbf{X}_{k}\}\subset\mathbb{X}\times\mathbb{I}$ be the globally labelled multi-object state estimate of node $a$ at time $k$ generated from the communicated message $(a,\mathbf{X}_{k})$, and $\mathbf{L}_{k}^{(a)}=\mathcal{L}(\mathbf{X}_{k}^{(a)})=\{\boldsymbol{\ell}\triangleq(\ell,a):\ell\in\mathcal{L}(\mathbf{X}_{k})\}\subset\mathbb{I}$ be its corresponding set of global labels.    The receiving node (\eg,~$b\in\mathcal{N}^{(a)}$) stores $\mathbf{X}_{k}^{(a)}$ in its own memory to form a set of globally labelled tracks $\mathbf{X}^{(a)}$ (or  $\mathbf{T}^{(a)}\subset\mathbb{T}\times\mathbb{I})$ from node $a$. Given the sets of globally labelled track estimates $\{\mathbf{T}^{(a)}\}_{a\in\mathcal{N}}$, our objective is to compute a consensed globally labelled multi-object state estimate $\mathbf{X}_{k}^{\text{con}}\subset\mathbb{X}\times\mathbb{I}$ by reaching a consensus on the latest labelled multi-object state estimates $\{\mathbf{X}_{k}^{(a)}\}_{a\in\mathcal{N}}$ in terms of kinematics and labels.  Moreover, we seek an efficient algorithm that is scalable with respect to (wrt) network size. A list of symbols is provided in Table~\ref{tab:parameters}.

Our solution is based on minimising the overall dissimilarity of tracks between sensor nodes. We employ a metric to measure such dissimilarities. Thus, in the next subsection, we revisit the metric property and a widely used metric in evaluating the tracking performance of MOT algorithms.

\subsection{Multi-object metrics}

A function $d:\mathcal{S}\times\mathcal{S}\rightarrow[0,\infty)$ is called a \textit{metric} or a \textit{distance} function on the space $\mathcal{S}$ if it meets the following conditions: 
\begin{itemize}
\item[1)] $d(x,y)=0$ if and only if $x=y$ (identity), 
\item[2)] $d(x,y)=d(y,x)$ (symmetry), 
\item[3)] $d(x,y)\leq d(x,z)+d(z,y)$ (triangle inequality). 
\end{itemize}
We are interested in distances between two finite subsets $X=\{x^{(1)},\dots,x^{(m)}\}$ and $Y=\{y^{(1)},\dots,y^{(n)}\}$ of a space equipped with a metric $d$, referred to as the base-distance (between the elements of $X$ and $Y$). An example of particular relevance to this work is the Optimal Sub-Pattern Assignment (OSPA) metric. 

Let $d^{(c)}(x,y)\triangleq\min(c,d(x,y))$, and $\Pi_{n}$ be the set of all permutations of $\left\{ 1,2,...,n\right\} $. The \textit{OSPA} metric of (integer) order $p\geq1$ and cut-off $c\in(0,\infty)$ is defined as ~\cite{schuhmacher2008consistent}:  
\begin{align}
 & d_{\mathtt{O}}^{(p,c)}(X,Y)\label{def:ospa_dist}\\
 & =\Bigg(\dfrac{1}{n}\bigg(\min_{\pi\in\Pi_{n}}\sum_{i=1}^{m}{d}^{(c)}(x^{(i)},y^{(\pi(i))})^{p}+c^{p}(n-m)\bigg)\Bigg)^{1/p},\nonumber 
\end{align}
if $m\leq n$, and $d_{\mathtt{O}}^{(p,c)}(X,Y)\triangleq d_{\mathtt{O}}^{(p,c)}(Y,X),$ if $m>n$, in addition, $d_{\mathtt{O}}^{(p,c)}(X,Y)=c$ if one of the argument is empty, and $d_{\mathtt{O}}^{(p,c)}(\emptyset,\emptyset)=0$. The two adjustable parameters $p$, and $c$, are interpreted as the  outlier and cardinality sensitivities, respectively. The OSPA metric
can be interpreted as the best-case localisation and cardinality error per object.

Given a meaningful base-distance $d^{(c)}(\cdot,\cdot)$ between two tracks, the  OSPA metric provides meaningful distances between two finite subsets of tracks. One such base distance is the time-averaged OSPA distance between the states of two tracks over time instances when at least one of the tracks exists \cite{beard2020a}.  Specifically, for any $t,u\in\mathbb{T}$, the \textit{OSPA track-to-track distance} is defined as

\begin{align}
\tilde{d}^{(c)}(t,u)=\sum\limits _{k\in\mathcal{D}^{(t)}\cup\mathcal{D}^{(u)}}\dfrac{d_{\mathtt{O}}^{(c)}(\{t(k)\},\{u(k)\})}{|\mathcal{D}^{(t)}\cup\mathcal{D}^{(u)}|}\label{def:ospa2_base}
\end{align}
if $\mathcal{D}^{(t)}\cup\mathcal{D}^{(u)}\neq\emptyset$, and $\tilde{d}^{(c)}(t,u)=0$, if $\mathcal{D}^{(t)}\cup\mathcal{D}^{(u)}=\emptyset$, where $d_{\mathtt{O}}^{(c)}(\{t(k)\},\{u(k)\})$ denotes the OSPA distance, and since the arguments are only sets of at most one element, the parameter $p$ no longer comes into play. For example, the OSPA track-to-track distance between the tracks $\mathbf{t}$ and $\mathbf{u}$ in Fig.~\ref{fig:track_def} is 
\begin{align}
    \tilde{d}^{(c)}(t,u) = 
    {\Big(7c + d_{\mathtt{O}}^{(c)}\big(\{t(k-4)\},\{u(k-4)\}\big) \Big) }/{8}. \notag
\end{align}

The OSPA metric with OSPA track-to-track distance in \eqref{def:ospa2_base}, called OSPA-on-OSPA or OSPA$^{(\text{2})}$, provides a natural distance between two sets of tracks. This distance can be interpreted as the time-averaged per-track error and demonstrates meaningful behaviour on various MOT scenarios. Errors in localisation, cardinality, track fragmentation and track identity switching all yield intended increases in the OSPA\textsuperscript{(2)} error. The higher the frequency of track fragmentation and identity switches, the higher the OSPA\textsuperscript{(2)} error. A dropped track that later regained with the same identity yields a smaller increase in OSPA\textsuperscript{(2)} error than if it were regained with a different identity ~\cite{beard2020a}.

\section{Fusion using Track Consensus}

\label{sec:info_fusion} This section presents a novel algorithm for fusing the latest local labelled multi-object state estimates among the distributed network of sensor nodes. Our innovation is based on the dissimilarity of tracks between nodes over multiple scans (\ie, time window) instead of relying on the multi-object densities at a single scan as in previous works. Fusing multi-object state estimates instead of multi-object densities significantly reduces the processing time and bandwidth. Moreover, the proposed solution applies to non-overlapping or partially overlapping limited FoVs of sensor nodes and improves tracking accuracy and cardinality estimations (or \textit{detectability} of objects).

We first consider the problem of measuring the dissimilarity between tracks. For this purpose, we define optimal track matching via the OSPA track-to-track distance. Based on the notion of optimal track matching, we formalise the concept of label consistency and derive a sufficient condition to achieve it. We then develop a novel algorithm for \textit{track consensus} by \textit{i)} formulating a method for achieving kinematic consensus between track estimates made by two nodes (Section~\ref{sec:kinematic_consensus_two}); \textit{ii)} extending this to a pair-wise fusion algorithm for kinematic consensus when the number of nodes is greater than two (Section~\ref{sec:kinematic_consensus_multi}); and \textit{iii)} developing a label consensus algorithm to achieve label consistency across the distributed network of nodes (Section~\ref{sec:achieving_label_consensus}).

\subsection{Optimal track matching}
Consider two nodes $a,b\in\mathcal{N}$. Without loss of generality, suppose that fusion is performed at node $a$. At time $k$, the aim is to compute a consensed labelled multi-object state estimate  $\mathbf{X}_{k}^{(a,\text{con})}$ at node $a$ from $\mathbf{X}_{k}^{(a)}$ and $\mathbf{X}_{k}^{(b)}$. Since   $\mathbf{X}_{k}^{(a,\text{con})}$ is computed from the latest local labelled multi-object state estimates, previously terminated tracks have no influence, only tracks declared to \textit{exist} at $k$ are considered. With a slight abuse of notation, let $\mathbf{T}_{k}^{(a)}=\{\mathbf{t}\in\mathbf{T}^{(a)}:\mathcal{L}(\mathbf{t})\in\mathcal{L}(\mathbf{X}_{k}^{(a)})\}$ be the set of live tracks at $k$, as declared by node $a$, truncated on the window $[j:k]$, and likewise for $\mathbf{T}_{k}^{(b)}=\{\mathbf{t}\in\mathbf{T}^{(b)}:\mathcal{L}(\mathbf{t})\in\mathcal{L}(\mathbf{X}_{k}^{(b)})\}$. For notational compactness, we use the shorthand $a^{(m)} = \mathcal{T}(\mathbf{a}^{(m)}) , b^{(m)} =  \mathcal{T}(\mathbf{b}^{(m)}),$ and $u^{(m)} =  \mathcal{T}(\mathbf{u}^{(m)})$.
\begin{definition}~\label{def:solution_of_optimal_matching}
Given two sets of tracks $\mathbf{T}_{k}^{(a)}=\{\mathbf{a}^{(1)},\dots,\mathbf{a}^{(|\mathbf{T}_{k}^{(a)}|)}\}$ and $\mathbf{T}_{k}^{(b)} = \{\mathbf{b}^{(1)},\dots,\mathbf{b}^{(|\mathbf{T}_{k}^{(b)}|)}\}$,  assuming that $|\mathbf{T}_{k}^{(a)}| \leq |\mathbf{T}_{k}^{(b)}|$. We define the optimal matching as the pairing of elements between the two sets that yields the OSPA\textsuperscript{(2)} distance $d_{\mathtt{O}}^{(1,c)}(\Tau(\mathbf{T}^{(a)}_k),\Tau(\mathbf{T}^{(b)}_k))$ ~\ie,
\begin{align}
\pi^{*}=\argmin_{\pi\in\Pi_{|\mathbf{T}^{(b)}_k|} }\sum_{m=1}^{|\mathbf{T}_{k}^{(a)}|}\tilde{d}^{(c)}(a^{(m)},b^{(\pi(m))}),\label{eq:optimal_matching}
\end{align}
Here, $\pi^*(m)=n$ means track $\mathbf{a}^{(m)}\in\mathbf{T}_{k}^{(a)}$ is matched to track $\mathbf{b}^{(n)}\in\mathbf{T}_{k}^{(b)}$. 
\end{definition}

The optimal assignment problem can be solved using the Hungarian algorithm~\cite{kuhn1955hungarian,munkres1957algorithms} with $\mathcal{O}(|\mathbf{T}^{(b)}_k|^{4})$ complexity, or using efficient algorithms in~\cite{tomizawa1971some,edmonds1972theoretical,jonker1987shortest} with $\mathcal{O}(|\mathbf{T}^{(b)}_k|^{3})$ complexity. 

\begin{remark}\label{remark:rank_assignment} The processing time of computing the optimal assignment in \eqref{eq:optimal_matching} is considerably shorter than that of LM-GCI in \cite{li2019computationally} where each node runs an LMB filter. The reason is that for the LMB filter, the multi-object state estimate is extracted from the LMB density for any label with existence probabilities higher than a predefined threshold (typically $0.5$). Thus, the number of tracks used in \eqref{eq:optimal_matching} is substantially smaller than the number of LMB components used in LM-GCI. Additionally, for optimal matching, we need to compute the distance between two tracks (for our method) or two LMB components (for LM-GCI). The computational complexity of computing OSPA\textsuperscript{(2)} distance between two tracks is considerably smaller than the computing distance (GCI-divergence) between two LMB components wherein a Gaussian mixture characterises the spatial density of each component.

\end{remark}

\subsection{Label consistency and performance bound}
In this subsection, we formalise the concept of label consistency for the consensus of labelled tracks. Note that while the term label consistency/inconsistency has been used in the context of DMOT in \cite{li2017robust}, there is no formal definition.  
\begin{definition}\label{def:label_consistency}
\textit{ Let $\mathbf{u}^{(1)},\dots,\mathbf{u}^{(N)} \in\mathbb{T}\times\mathbb{I}$ be the true trajectories of $N$ distinct objects. Let $\mathbf{a}^{(1)},\dots,\mathbf{a}^{(N)}$ be their corresponding track estimates at node $a$,  and $\mathbf{b}^{(1)},\dots,\mathbf{b}^{(N)}$ be their corresponding track estimates at node $b$. Then, we say that label consistency is achievable when the optimal matching $\pi^{*}$ between these sets of track estimates satisfies 
\begin{align}
\pi^*(m)=m,~\forall m\in\{1,\dots,N\}.\label{eq:H_star_mm}
\end{align}
Moreover, suppose that for each $m\in\{1,\dots,N\}$ the tracks $\mathbf{a}^{(m)}$ and $\mathbf{b}^{(m)}$ are assigned new labels, resulting in tracks $\bar{\mathbf{a}}^{(m)}$ and $\bar{\mathbf{b}}^{(m)}$. Then, we say label consistency is achieved if
\begin{align}
\mathcal{L}(\bar{\mathbf{a}}^{(m)})=\mathcal{L}(\bar{\mathbf{b}}^{(m)}),~\forall m\in\{1,\dots,N\}.
\end{align}
}\end{definition} 
In essence, label consistency is achieved if the same object is assigned the same global label by two nodes. Otherwise, we have label inconsistency.

\begin{remark}~\label{remark:label_consistency} Definition~\ref{def:label_consistency}
can be easily extended to more than two nodes. Suppose that $\bar{\mathbf{t}}^{(1)},\dots,\bar{\mathbf{t}}^{(|\mathcal{N}|)}\in\mathbb{T}\times\mathbb{I}$ are the globally labelled track estimates of a true trajectory $\mathbf{u}$. Then label consistency is achieved if 
\begin{align}
\mathcal{L}(\bar{\mathbf{t}}^{(m)})=\mathcal{L}(\bar{\mathbf{t}}^{(n)}),~\forall m,n\in\{1,\dots,|\mathcal{N}|\}.
\end{align}
\end{remark}
The following result provides the conditions to achieve label consistency (see Appendix~\ref{sec:math_proofs} for proofs).
\begin{proposition}\label{lem:dist_bound} Consider the trajectories $\mathbf{u}^{(1)},\dots,\mathbf{u}^{(N)}$ of $N\leq\min(|\mathbf{T}_{k}^{(a)}|,|\mathbf{T}_{k}^{(b)}|)$ distinct objects where $\mathbf{T}_{k}^{(a)}$ and $\mathbf{T}_{k}^{(b)}$ are the sets of live tracks at time $k$ at node $a$ and node $b$, respectively. Let us denote their corresponding labelled track estimates at nodes $a$ and $b$, by $\mathbf{a}^{(1)},\dots,\mathbf{a}^{(N)}$ and $\mathbf{b}^{(1)},\dots,\mathbf{b}^{(N)}$.  Suppose that their track-to-track errors $\tilde{d}^{(c)}(a^{(m)},u^{(m)})$, $\tilde{d}^{(c)}(u^{(m)},b^{(m)})$ at nodes $a$ and $b$, are bounded by $\mathcal{E}$ for all $m\in\{1,\dots,N\}$, and that: 
\begin{equation}
\begin{gathered}\tilde{d}^{(c)}(a^{(m)},b^{(n)})=c~\forall m>N,\forall n\in\{1,\dots,|\mathbf{T}_{k}^{(b)}|\},\\
\tilde{d}^{(c)}(a^{(m)},b^{(n)})=c~\forall m\in\{1,\dots,|\mathbf{T}_{k}^{(a)}|\},\forall n>N.
\end{gathered}
\label{eq:remaing_tracks_dist}
\end{equation}
Then, to achieve label consistency for these $N$ distinct objects, we need 
\begin{align}
\tilde{d}^{(c)}(u^{(m)},u^{(n)})>4\mathcal{E}~\forall m,n\in\{1,\dots,N\},m\neq n.\label{eq:label_consistent_condi}
\end{align}
\end{proposition} 
For a given upper bound on the track-to-track error, the above result provides a lower bound on the separation of the true trajectories required to achieve label consistency. A conservative track-to-track error bound can be determined from the kinematic error and the frequency of misses or the empirical existence probability.  
\begin{definition}\label{def:pos_px} The empirical existence probability of a track $t$ over the time interval $\{j:k\}$ is defined as 
\begin{align}
P_{X}(t)=\dfrac{|\mathcal{D}^{(t)}|}{k-j+1},
\end{align}
where $\mathcal{D}^{(t)}\subseteq\{j:k\}$ is the domain of track $t$. 
\end{definition}
\begin{proposition} Let $P_{X}^{\min}$ be minimum empirical existence probabilities of all tracks at nodes $a$ and $b$, ~\ie, 
\begin{align}
P_{X}^{\min}=\min_{\mathbf{t}\in\mathbf{T}_{k}^{(a)}\cup\mathbf{T}_{k}^{(b)}}P_{X}(\mathcal{T}(\mathbf{t})),
\end{align}
Suppose the single-scan estimation errors at nodes $a$ and $b$ are bounded as follows: 
\begin{equation}
\begin{gathered}d_{\mathtt{O}}^{(c)}(\{a^{(m)}(i)\},\{u^{(m)}(i)\})=\begin{cases}
c & i\notin\mathcal{D}^{(\mathbf{a}^{(m)})}\\
\notag d^{(c)}(a^{(m)}(i),u^{(m)}(i))\leq\varepsilon & i\in\mathcal{D}^{(\mathbf{a}^{(m)})}
\end{cases},\\
d_{\mathtt{O}}^{(c)}(\{b^{(m)}(i)\},\{u^{(m)}(i)\})=\begin{cases}
c & i\notin\mathcal{D}^{(\mathbf{b}^{(m)})}\\
d^{(c)}(b^{(m)}(i),u^{(m)}(i))\leq\varepsilon & i\in\mathcal{D}^{(\mathbf{b}^{(m)})}
\end{cases}, \\
\forall m\in\{1,\dots,N\}.
\end{gathered}
\end{equation}
Then the track-to-track error is bounded by
\begin{align}
    \varepsilon P_{X}^{\min}+c(1-P_{X}^{\min}).
\end{align}
\end{proposition} \begin{remark}\label{remark:pD} If the window $[j:k]$ is long enough, and a ``good'' multi-object tracker (\eg,
MHT~\cite{reid1979algorithm}, GLMB~\cite{vo2013glmb,vo2014glmb} or multi-scan GLMB~\cite{vo2019multiscan}) is used, then it is possible to have $P_{X}^{\min}\geq P_D$ where $P_D$ is the detection probability. Thus:
\begin{align}
\varepsilon P_{X}^{\min}+c(1-P_{X}^{\min})\leq \varepsilon P_{D}+c(1-P_{D}),
\end{align}
and the condition for label consistency is 
\begin{align} \label{eq:label_consistency_result}
\tilde{d}^{(c)}(u^{(m)},u^{(n)})&>4\big[\varepsilon P_{D}+c(1-P_{D})\big],\notag\\ 
\forall m,n&\in\{1,\dots,N\},m\neq n. 
\end{align}
Intuitively, \eqref{eq:label_consistency_result} suggests that to achieve label consistency it is necessary that \textit{i)}~the object true trajectories are reasonably well-separated; \textit{ii)} the detection probability $P_{D}$ is relatively high.
\end{remark} 

\subsection{Kinematic consensus for two nodes} \label{sec:kinematic_consensus_two}
\textit{Track consensus} involves kinematic consensus and label consensus. In this subsection, we develop an algorithm to compute the \textit{kinematic consensus} between track estimates from two nodes. This kinematic consensus algorithm is extended to multiple nodes in Section~\ref{sec:kinematic_consensus_multi}, while the label consensus algorithm is discussed in Section~\ref{sec:achieving_label_consensus}.

Since each node is equipped with a limited FoV sensor, only track estimates within the overlapped FoV of two sensors (observed\footnote{When we say track $t$ is observed at node $a$ means that node $a$ declares track $t$ exist.} by both nodes' FoVs) should be matched and fused to enhance tracking accuracy. Further, the unmatched tracks (outside the overlapped FoV of two sensors) should be retained to improve detectability. Thus, the set of tracks at each node can be divided into three subsets: \textit{i)} matched tracks, \textit{ii)} unmatched and retained tracks, \textit{iii)} unmatched and discarded tracks. However, for distributed sensor networks, the sensor FoV information of other nodes is usually unknown, and we rely on the separation between two-track estimates to determine if they are matched or not. For all pairs $(m,n)$ such that $\pi^*(m)=n$, only pairs with associated costs $C_{m,n} = \tilde{d}^{(c)}(a^{(m)},b^{(n)})$  less than $c$ are considered matched. Essentially, this implies that both matched tracks must exist together for at least one instance (\eg, at time $k-4$ as in Fig.~\ref{fig:track_def}) or are not too distant in accordance with the cut-off distance $c$. 
\begin{align}
m\in Q_{k}^{(a)} & \subseteq\{1,\dots,|\mathbf{T}_{k}^{(a)}|\},\nonumber \\
n\in Q_{k}^{(b)} & \subseteq\{1,\dots,|\mathbf{T}_{k}^{(b)}|\},\label{eq:mn_in_Q}\\
\text{ subject to } & \pi^*(m)=n,C_{m,n}<c.\nonumber 
\end{align}
The steps for determining matched pairs are given in Algorithm~\ref{Algo_DetermineMatchedPairs}
(see Appendix~\ref{sec:pseudo_codes}). 

For the unmatched tracks, some of these can be false tracks and should be discarded. Thus, we propose that only unmatched tracks with lengths higher than a predefined track length $\mathcal{C}_{\text{len}}$ are retained. Similar to the traditional thresholding method in extracting raw measurements, a small track length $\mathcal{C}_{\text{len}}$ helps to initiate new objects faster albeit with more false-positives and vice versa~\cite{ristic2013tutorial}. Intuitively, $\mathcal{C}_{\text{len}}$ delays the confirmation of a new object by the track consensus algorithm for at least  $\mathcal{C}_{\text{len}}$ steps, \ie, the observed track must have a track length greater than or equal to $\mathcal{C}_{\text{len}}$ to be confirmed as a new birth object. As a result, the consensed labelled multi-object state estimate comprises of two components: \textit{i)} the matched and fused labelled state estimates, \textit{ii)} the unmatched and retained labelled state estimates of two nodes. The steps for fusing two nodes are described in Algorithm~\ref{Algo_FuseTwoNodes}, and elaborated on below:

\vspace{1mm}
\noindent\textit{Step 1) Fuse matched labelled state estimates:} We compute\footnote{We implement a naive solution here such that the fused label is the label of the node performing the fusion steps while further improvements to achieve label consensus are discussed in Section~\ref{sec:achieving_label_consensus}. Here, we employ a simple weighted average of the local estimates,  } the set $\bar{\mathbf{X}}_{k}^{(a)}$ of matched labelled state estimates at node $a$ by fusing matched labelled state estimates,~\ie,
\begin{gather}\label{eq:X_fused}
    \bar{\mathbf{X}}_{k}^{(a)} = \big\{ (x,\boldsymbol{\ell}^{(m)}): x = w^{(a)}a^{(m)}(k)+w^{(b)}b^{(n)}(k), (m,n) \in Q_k \big\},
\end{gather}
where $w^{(a)}$ and $w^{(b)}$ be the fusing weights of two nodes with $w^{(a)}+w^{(b)}=1$ and $w^{(a)},w^{(b)}>0$ (See \eqref{eq:compute_fusing_weight} later in Section~\ref{sec:experiments} for how the weights are selected). Notably, we also maintain an  $|\mathbf{L}_{1:k}^{(a)}|\times|\mathbf{L}_{1:k}^{(b)}|$ \textit{matched history} matrix $\Xi_{1:k}^{(a,b)}$ (line $4$ in Algorithm~\ref{Algo_FuseTwoNodes})  whose elements are the number of instances that track $i^{(m)}\in\{1,\dots,|\mathbf{L}_{1:k}^{(a)}|\}$ is matched with track $i^{(n)}\in\{1,\dots,|\mathbf{L}_{1:k}^{(b)}|\}$. Here, $\mathbf{L}_{1:k}^{(a)}$ is the label space up to time $k$ at node $a$ (likewise for $\mathbf{L}_{1:k}^{(b)}$ at node $b$).  We describe in Algorithm~\ref{Algo_UpdateMatchedHistory} in Appendix~\ref{sec:pseudo_codes} the method for updating the matched history matrix.

\vspace{1mm}
\noindent\textit{Step 2) Retain unmatched labelled state estimates:} 
We retain the set $\mathbf{X}_{k}^{(a,\text{ret})}$ of unmatched labelled state estimates at node $a$ (likewise for $\mathbf{X}_{k}^{(b,\text{ret})}$ at node $b$) whose lengths exceed a given $\mathcal{C}_{\text{len}}$, \ie,~
\begin{gather}\label{eq:X_ret}
    \mathbf{X}_{k}^{(a,\text{ret})} = \big\{ \mathbf{a}^{(m)}(k):  m\in\{1,\dots,|\mathbf{T}_{k}^{(a)}|\}\setminus Q_{k}^{(a)}, |\mathcal{D}^{(\mathbf{a}^{(m)})}|\geq\mathcal{C}_{\text{len}} \big\}.
\end{gather}
Further, any unmatched tracks with lengths less than $\mathcal{C}_{\text{len}}$ are discarded. 

\noindent\textit{3) Compute consensed labelled state estimates:} We compute the set $\mathbf{X}_{k}^{(a,\text{con})}$ of consensed labelled state estimates at node $a$ by combining two components: \textit{i)} the matched and fused labelled state estimates, \textit{ii)} the unmatched and retained labelled state estimates of two nodes, \ie,
\begin{align}
\mathbf{X}_{k}^{(a,\text{con})}=\bar{\mathbf{X}}_{k}^{(a)}\cup\mathbf{X}_{k}^{(a,\text{ret})}\cup\mathbf{X}_{k}^{(b,\text{ret})}.\label{eq:report_fused_estimate}
\end{align}

\noindent \begin{remark}\label{remark:consensed_est} The consensed
labelled multi-object state estimate $\mathbf{X}_{k}^{(a,\text{con})}$ does not play a role in optimal matching at the next time step and is used for reporting purposes only. The local multi-object state estimate $\mathbf{X}_{k+1}^{(a)}$ at the next time step $k+1$ is independent of $\mathbf{X}_{k}^{(a,\text{con})}$. The optimal matching at $k+1$ only depends on sets of live tracks $\mathbf{T}^{(a)}_{k+1}$ and $\mathbf{T}^{(b)}_{k+1}$, not on the consensed multi-object state estimate $\mathbf{X}_{k}^{(a,\text{con})}$.
\end{remark}

\begin{algorithm}[!tb]
	\footnotesize
	\caption{\textsf{FuseTwoNodes}}\label{Algo_FuseTwoNodes}
	\begin{algorithmic}[1] 
		\Statex \textbf{Input}: $~\mathbf{T}_k^{(a)};~\mathbf{T}_k^{(b)};~\Xi^{(a,b)}_{1:k-1};~\mathbf{L}^{(a)}_{1:k-1};~\mathbf{L}^{(b)}_{1:k-1};~\mathcal{C}_{\text{len}};$
		\Statex \textbf{Output}: $~\mathbf{X}^{(a,\text{con})}_k;~\Xi^{(a,b)}_{1:k};~\mathbf{L}^{(a)}_{1:k};~\mathbf{L}^{(b)}_{1:k};$
		\State $\mathbf{L}^{(a)}_{1:k} = \mathbf{L}^{(a)}_{1:k-1} \cup \big(\mathcal{L}(\mathbf{T}^{(a)}_k) \setminus \mathbf{L}^{(a)}_{1:k-1} \big); $
		\State $\mathbf{L}^{(b)}_{1:k} = \mathbf{L}^{(b)}_{1:k-1} \cup \big(\mathcal{L}(\mathbf{T}^{(b)}_k) \setminus \mathbf{L}^{(b)}_{1:k-1} \big); $		
		\State $Q_k := \textsf{\footnotesize {DetermineMatchedPairs}}(\mathbf{T}_k^{(a)},\mathbf{T}_k^{(b)}));$
		\State $\Xi^{(a,b)}_{1:k} := \textsf{\footnotesize {UpdateMatchedHistory}}(\Xi^{(a,b)}_{1:k-1},~\mathbf{L}^{(a)}_{1:k},~\mathbf{L}^{(b)}_{1:k},~Q_k);$		
		\State \textsf{\footnotesize {Compute }} $\bar{\mathbf{X}}^{(a)}_{k}$ \textsf{\footnotesize {~via~}} \eqref{eq:X_fused}  $;$ $\mathbf{X}^{(a,\text{ret})}_{k}$ and $\mathbf{X}^{(b,\text{ret})}_{k}$\textsf{\footnotesize {~via~}} \eqref{eq:X_ret} $;$ 
		\State $\mathbf{X}^{(a,\text{con})}_k := \bar{\mathbf{X}}^{(a)}_k \cup  \mathbf{X}^{(a,\text{ret})}_{k} \cup \mathbf{X}^{(b,\text{ret})}_{k}; $ 
	\end{algorithmic}
\end{algorithm}

\subsection{Kinematic consensus for multiple nodes} \label{sec:kinematic_consensus_multi}
The previous subsections presented a new kinematic consensus algorithm for fusing multi-object state estimates between two nodes. We now extend the algorithm to more than two nodes. A direct extension results in an $|\mathcal{N}|$-dimensional optimal assignment for the track matching, an NP-hard problem. A widely used sub-optimal strategy to circumvent this intractable problem is to perform pair-wise matching sequentially~\cite{battistelli2013consensus,fantacci2015consensus,li2019computationally}.  Based on the proposed network architecture in Section~\ref{sec:problem_description}, the kinematic consensus is realised in Algorithm~\ref{Algo_FuseMultiNodes}, called TC-OSPA\textsuperscript{(2)}. Algorithm~\ref{Algo_FuseMultiNodes} consists of two steps: 

\noindent\textit{Step 1) Fuse each neighbour node with current node:} Each neighbour labelled multi-object state estimate is fused to the labelled multi-object state estimate of the node of interest to ensure that only unmatched labelled state estimates with the track lengths greater than or equal to $\mathcal{C}_{\text{len}}$ are retained.

\noindent\textit{Step 2) Fuse the consensed labelled state estimates sequentially: }  The consensed labelled state estimates of neighbouring nodes and the nodes of interest are combined sequentially without the track length constraint (\ie,~ $\mathcal{C}_{\text{len}}=1$) to retain all unmatched tracks because the label consensus has not been reached.
\begin{algorithm}[!ht]
	\footnotesize
	\caption{\textsf{FuseMultiNodes}}\label{Algo_FuseMultiNodes}
	\begin{algorithmic}[1] 
		\Statex \textbf{Input}: $~\{\mathbf{T}_k^{(a)}\}_{a \in \mathcal{N} };~\{\Xi^{(a,b)}_{1:k-1}\}_{a,b \in \mathcal{N}};~\{\mathbf{L}^{(a)}_{1:k-1}\}_{a \in \mathcal{N} };$
		\Statex \textbf{Output}: $~\{\mathbf{X}^{(a,\text{con})}_{k}\}_{a \in \mathcal{N} };~~\{\Xi^{(a,b)}_{1:k}\}_{a,b \in \mathcal{N}};~\{\mathbf{L}^{(a)}_{1:k}\}_{a \in \mathcal{N} };$
		\For {$a=1:|\mathcal{N}|$}
		\State $\mathcal{B}:=\{1:|\mathcal{N}|\} \setminus \{a\}; $
		\State $\mathbf{X}_{\text{temp}} = [];$ \Comment{Initialise a temporarily consensed   ~~~~~~~~~~~~\quad\quad\quad\quad\quad\quad\quad\quad\quad\quad\quad labelled multi-object state estimate.}
		\For{$i=1:|\mathcal{B}|$} 
		\Statex{\quad \quad \quad $\triangleright$ Step 1:  fuse each neighbour node with current node}.
		\State $b := \mathcal{B}(i);$
		\State $[\mathbf{X}_{\text{temp}}^{(i)},\Xi^{(a,b)}_{1:k},\mathbf{L}^{(a)}_{1:k},\mathbf{L}^{(b)}_{1:k}]:=$
		\State \textsf{\footnotesize{FuseTwoNodes}}$(\mathbf{T}_k^{(a)},\mathbf{T}_k^{(b)},\Xi^{(a,b)}_{1:k-1},\mathbf{L}^{(a)}_{1:k-1},\mathbf{L}^{(b)}_{1:k-1},\mathcal{C}_{\text{len}})$; 
		\EndFor
		\State $\mathbf{X}^{(a,\text{con})}_{k} :=\mathbf{X}_{\text{temp}}^{(1)};$
		\If{$|\mathcal{B}| > 1 $} 
		\Statex{\quad \quad \quad $\triangleright$ Step 2: fuse the consensed labelled state estimates sequentially.} 
		\For{$i=2:|\mathcal{B}|$}
		\State $\mathbf{X}^{(a,\text{con})}_{k}:=$\textsf{\footnotesize{FuseTwoNodes}}$(\mathbf{X}^{(a,\text{con})}_{k},\mathbf{X}_{\text{temp}}^{(i)},[],[],[],1)$; 
		\EndFor
		\EndIf
		\State $\mathbf{L}^{\text{con}}_{a,k} := \textsf{\footnotesize {UpdateLabels}}(\mathcal{L}(\mathbf{X}^{(a,\text{con})}_{k});\{\Xi^{(a,b)}_{1:k}\}_{a,b \in \mathcal{N}};\{\mathbf{L}^{(a)}_{1:k}\}_{a \in \mathcal{N} });$
		\EndFor
	\end{algorithmic}
\end{algorithm}

Algorithm~\ref{Algo_FuseMultiNodes} (TC-OSPA\textsuperscript{(2)}) can be implemented in real-time given the significantly low message sizes realised by transmitting labelled multi-object state estimates compared to transmitting labelled multi-object densities. In particular, suppose that  $|\mathbf{T}_{\max}|$ is the maximum number of objects seen by the network, \ie,~$|\mathbf{T}_{\max}|=\max\big(|\mathbf{T}^{(1)}|,\dots,|\mathbf{T}^{(|\mathcal{N}|)}|\big)$, then the order of magnitude of data that needs to be shared is upper bounded by\footnote{The upper bound is only reached in the case that all nodes observe all the objects. In reality, because of limited FoV sensors, this upper bound will not be reached in most cases.} 
\begin{align}
|\mathcal{N}||\mathbf{T}_{\max}|.
\end{align}
For example, if $|\mathcal{N}|=20$ nodes, $|\mathbf{T}_{\max}|=100$ objects, and each labelled state estimate has $6$ dimensions ($4$ for 2D environments of kinematic state and $2$ for the local label's dimension), and each dimension is represented by an 8-byte floating-point value. The maximum amount of data that needs to be shared by a node at one time is $93.75$~KB, which is reasonably low to track a large number of objects using $20$ distributed nodes. 
Notably, our method does not require transmitting all the trajectories because the state estimates from other nodes can be stored in each local node. Even if the trajectories need to be sent, for instance, due to limited memory at each local node, the proposed method only needs to transmit truncated trajectories (\eg, $10$ scans). Consequently, the bandwidth needed is much smaller than sending a multi-object density and is proportional to the number of objects. 
Further, the computational complexity of Algorithm~\ref{Algo_FuseMultiNodes} at each local node is $\mathcal{O}(|\mathcal{N}||\mathbf{T}_{\max}|^4 d)$ where $d$ is the dimension of a single-object state. Thus, the computational complexity only increases linearly with respect to the number of nodes.  See numerical experiments in Sec.~\ref{sec:experiments} for fusing time.
\begin{remark}
Since our approach is agnostic to local multi-object tracking techniques, we do not assume that the nodes also provide covariances of their estimates. Nonetheless, if the local trackers can provide covariances, then our approach is also applicable by fusing the means and covariances. In particular, we can: i)~compute the distance in \eqref{def:ospa2_base} using the Mahalanobis distance between two Gaussian distributions;  ii)~compute  the fused estimate in \eqref{eq:X_fused}  and its covariance  using methods such as the optimal fusion method for two-sensors~\cite{barshalom1986the} or GCI for multiple sensors.
\end{remark}

\begin{figure*}[!tb]
\centering \includegraphics[width=0.8\textwidth]{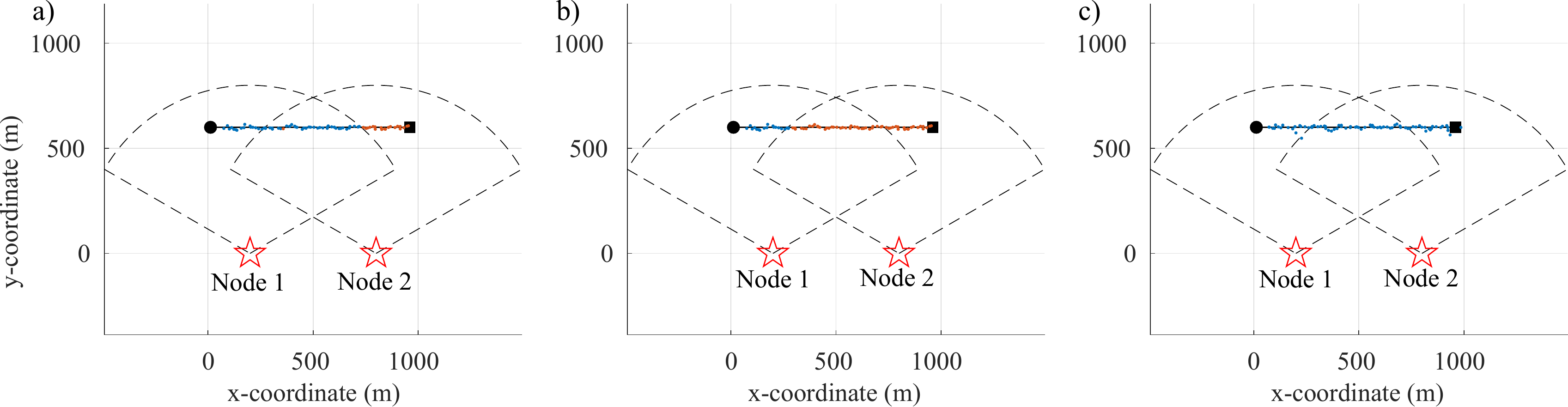} \vspace{-0.2cm}
\caption{Example~\ref{example:1} --- two nodes with limited FoVs tasked with tracking objects. a) Consensed labelled state estimates at node $1$ before reaching label consensus; b) Consensed labelled  state estimates at node $2$ before reaching label consensus; c) Consensed labelled state estimates at node $1$ (likewise for node $2$) after reaching label consensus. Here: `$\circ$' is the location of the object's birth; `$\Box$' is the location of the object's death. Colour coding represents labels of the objects. }
\vspace{-0.5cm}
 \label{fig:Ex1} 
\end{figure*}

\subsection{Label consensus for multiple objects} ~\label{sec:achieving_label_consensus} 
The previous subsections address multiple limited FoVs \textit{kinematic consensus}. However, MOT concerns not only the object's positions (kinematic) but also the object's identities (labels). In this subsection, we present our \textit{label consensus} solution that reduces the occurrences of label inconsistency. During the fusion steps, the unmatched  and retained labelled state estimates are included in the consensed labelled multi-object state estimates (see \eqref{eq:report_fused_estimate}). Hence, care must be taken to ensure mismatched labels between nodes are resolved to achieve \textit{label consensus}, as illustrated in the following example.

\begin{example}
\label{example:1} Consider using a two-node distributed sensor network to track a mobile object that follows a constant velocity model over a $[-500,1500]$m $\times[0,1000]$m area. The locations of the two nodes are $[0,400]^{T}$~m and $[0,800]^{T}$~m, respectively. Each node runs an LMB filter locally, and its sensor can only detect objects within its relative angle $[-60^{\circ},60^{\circ}]$ with detection probability $P_{D}=0.98$ and detection range $r_{D}=800$~m (see Fig.~\ref{fig:Ex1}).  After the kinematic consensus steps, we have the two following label inconsistency problems:

\vspace{1mm}
\noindent\textit{1)~Label inconsistency of the unmatched and retained labelled state estimates.~} Fig.~\ref{fig:Ex1}a depicts the consensed labelled  state estimates \textit{at node $1$}. Although the kinematic fusion successfully helps node $1$ track the object, a track fragmentation still occurs even when it moves out of node 1's FoV.  This happens because it is no longer detected by node $1$, and the network relies on detection information from node $2$ (including node $2$'s label) to track it. Since node $2$ has a different label (in red) to the one observed by node $1$ (in blue), the consensed labelled state estimates yield two different labels for the same object.

\vspace{1mm}
\noindent\textit{2)~Label inconsistency of the matched labelled state estimates.~} Fig.~\ref{fig:Ex1}b depicts the consensed labelled state estimates \textit{at node $2$}. Initially, the object is not detected by node $2$; hence, the network relies on the information from node $1$ (including node $1$'s label in blue) to track it. When the object moves into node $2$'s FoV, node $2$ assigns it a new label in red. During the fusion steps (see \eqref{eq:X_fused}), the object is assigned the red label of node $2$ (since the fusion is performed at node $2$). As a result, the consensed labelled state estimates have different labels for the same object. 
\end{example}

In the following, we present a label consensus solution to reduce label inconsistency.  The main idea is to construct a graph that represents connections among labels based on the match history matrices $\Xi_{1:k}^{(a,b)}$ whose $(m,n)$ entry is the number of instances the $m$th track in node $a$'s track-list is matched with the $n$th track in node $b$'s track-list. In particular, let $G=(V, E)$ be a graph, where: 
\begin{itemize}
\item the set $V$ of labels is the set of vertices of the graph;
\item $E$ is the set of edges representing matches between labels. 
\end{itemize}
Since  label consistency requires the same object to have the same label across all nodes of the network (Remark~\ref{remark:label_consistency}), all of the connected vertices should be assigned the same label. For this so-called consensed label, we propose to use the least label of connected vertices according to a lexicographical order of time of birth and unique node ID : 
\begin{align}
 & \boldsymbol{\ell}=((k,i),a)<\boldsymbol{\ell}'=((k',i'),b)\nonumber \\
\Leftrightarrow~ & (k=k')\text{ and }(a<b)\text{ or }(k<k').\label{eq:label_comparison}
\end{align}
The proposed label consensus is provided in Algorithm~\ref{Algo_UpdateLabels} in Appendix~\ref{sec:pseudo_codes}, which is used in line $15$ of Algorithm~\ref{Algo_FuseMultiNodes}. Fig.~\ref{fig:Ex1}c depicts the consensed labelled multi-object state estimate at node $1$ after label consensus is achieved. Observe that label inconsistency is eliminated.

\section{Numerical Experiments}

\label{sec:experiments} In this section, we apply the proposed TC-OSPA\textsuperscript{(2)} fusion method\footnote{see \textit{\url{https://github.com/AdelaideAuto-IDLab/Distributed-limitedFoV-MOT}} for our open source code release with detailed implementation.} to investigate and compare with other fusion strategies in three distributed sensor network settings of increasing complexity.  A 2-dimensional search area is adopted for these three scenarios to
demonstrate the effectiveness of our method. Each detected object with  kinematic state $x=[p_{x},\dot{p}_{x},p_{y},\dot{p}_{y}]^{T}$ results in an observation $z$ of noisy $xy$-coordinate positions, given by: $z=\big[p_{x},p_{y}]^{T}+v$. Here, $v\sim N(0,R)$ is a $2\times1$ zero-mean Gaussian process
noise with $R=\mathrm{diag}(\sigma_{x}^{2},\sigma_{y}^{2})$ where $\sigma_{x}=\sigma_{y}=10$~m.  Each object follows a constant velocity model given by $x_{k}=F^{CV}x_{k-1}+q_{k-1}^{CV}$. Here, $F^{CV}=[1,T_{0};0,T_{0}]\otimes I_{2}$, $T_{0}$ is the sampling interval ($T_{0}=1$~s for our experiments), $\otimes$ denotes for the Kronecker tensor product; $I_{2}$ is the $2\times2$ identity matrix; $q_{k-1}^{CV}\sim N(0,Q^{CV})$ is a $4\times1$ zero-mean Gaussian process noise, with co-variance $Q^{CV}=\sigma_{CV}^{2}[T_{0}^{3}/3,T_{0}^{2}/2;T_{0}^{2}/2,T_{0}]\otimes I_{2}$ where $\sigma_{CV}=5$~m/s\textsuperscript{2}.  Each object has survival probability $P_{S}=0.98$. Clutter follows a Poisson model with an average of $10$ clutters per scan. We use OSPA and OSPA\textsuperscript{(2)} as well as fusing time to measure performance. For OSPA and OSPA\textsuperscript{(2)}, we set  cut-off $c=100$~m, order $p=1$. The OSPA\textsuperscript{(2)} distance at time $k$ is calculated over a 10-scan window ending at $k$. The fusing time reported is the average execution time for the fusion steps over 100 MC runs---notably, this excludes the execution of the filtering algorithm. 

For a fair comparison with LM-GCI in~\cite{li2019computationally}, we use an LMB filter at each local node, \textit{although our approach can be used with different filters} such as MHT or GLMB since our method is agnostic to the filters at each node. The LMB filter is implemented with Gaussian mixtures using Gibbs sampling~\cite{vo2016efficient} for a joint prediction and update step. The existence threshold is set at $10^{-3}$, \ie,~any Bernoulli component with label $\ell$ and its existence probability $r^{(\ell)}<10^{-3}$ is pruned. A Bernoulli component with label $\ell$ with existence probability $r^{(\ell)}>0.5$ is confirmed as an existing object, and its state estimates are extracted from the corresponding state density. Further, we implement the Adaptive Birth Procedure (ABP) in~\cite{reuter2014lmb}. In particular, the birth distribution $\mathbf{\pib}_{B,k+1}$ at time $k+1$ is a function of measurement sets $Z_{k}$, \ie,~$\mathbf{\pib}_{B,k+1}=\big\{ r_{B,k+1}^{(\ell)}(z),p_{B,k+1}^{(\ell)}(x|z)\big\}_{l=1}^{|Z_{k}|}$, where 
\begin{align*}
r_{B,k+1}^{(\ell)}(z)=\min\Big(r_{B,\max},\dfrac{1-r_{U,k}(z)}{\sum_{\zeta\in Z_{k}}1-r_{U,k}(\zeta)}\lambda_{B,k+1}\Big).
\end{align*}
Here, $r_{U,k}(z)$ is the probability that the measurement $z$ is associated with tracking hypotheses given by 
\begin{align}
r_{U,k}(z)=\sum_{I_{k-1},\xi,I_{k},\theta_{k}}1_{\theta_{k}}(z)w^{(I_{k-1},\xi)}w^{(I_{k-1},\xi,I_{k},\theta_{k})},
\end{align}
where $w^{(I_{k-1},\xi)}w^{(I_{k-1},\xi,I_{k},\theta_{k})}$ is in~\cite[eq.14]{vo2016efficient}, $\lambda_{B,k+1}$ is the expected number of births at time $k+1$ and $r_{B,\max}$ is the maximum existence probability of a newly born object. In three scenarios, we set $\lambda_{B,k+1}=0.5$ and $r_{B,\max}=0.03$.

For data fusion, since we do not focus on the weight selection problem, the Metropolis weight~\cite{xiao2005scheme} is implemented, \ie,~for each node  $a,b\in\mathcal{N}$, 
\begin{align} \label{eq:compute_fusing_weight} 
w^{(a,b)}=\begin{cases}
\frac{1}{1+\max(|\mathcal{N}^{(a)}|,|\mathcal{N}^{(b)}|)}, & a\in\mathcal{N},b\in\mathcal{N}^{(a)},\\
1-\sum_{b\in\mathcal{N}^{(a)}}w^{(a,b)}, & a\in\mathcal{N},b=a.
\end{cases}
\end{align}

\begin{figure*}[!tb]
\centering \includegraphics[width=1.0\textwidth]{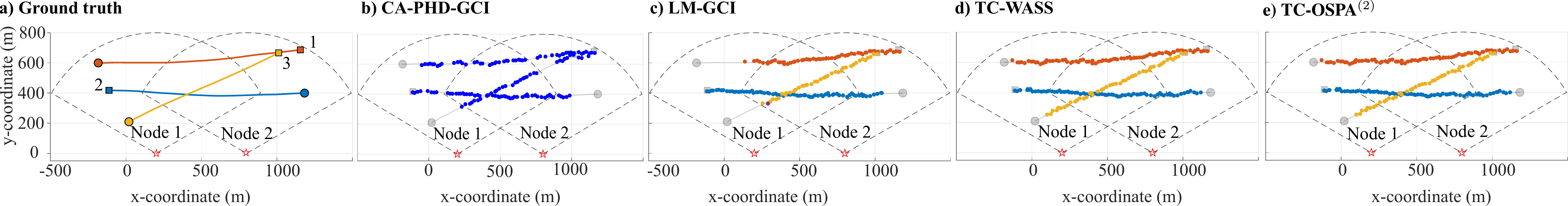}
\vspace{-0.6cm}
 \caption{Scenario 1 ground-truth and consensed labelled multi-object state estimates at sensor node $2$: a) ground-truth; b) CA-PHD-GCI (unlabelled approach); c) LM-GCI; d) TC-WASS; e) TC-OSPA\protect\textsuperscript{(2)}. Starting and stopping positions are denoted by $\circ$ and $\Box$, respectively. Colour coding represents objects' labels. 
 }
\vspace{-0.6cm}
 \label{fig:Scenario10_est_vs_truth_settings} 
\end{figure*}

We compare the proposed fusion method TC-OSPA\textsuperscript{(2)}, with CA-PHD-GCI~\cite{li2020Guchong} and LM-GCI~\cite{li2019computationally}. We also compare our proposed approach with the OSPA metric replaced by the \textit{Wasserstein} metric,  here on referred to as TC-WASS. The Wasserstein metric of (integer) order $p\geq1$ is defined
as~\cite{hoffman2004multitarget}: 
\begin{align}\label{eq:wasserstein_dist}
d_{\mathtt{W}}^{(p)}(X,Y) & \triangleq\min_{\bar{C}}\Big(\sum_{i=1}^{m}\sum_{j=1}^{n}\bar{C}_{i,j}d(x^{(i)},y^{(j)})^{p}\Big)^{1/p}
\end{align}
where $\bar{C}=\left(\bar{C}_{i,j}\right)$ denotes an $m\times n$ transportation matrix, \ie,~ the entries $\bar{C}_{i,j}$ are non-negative, each row sum to $1/m$, and each column sum to $1/n$. The integer $p$ determines the sensitivity of the metric to outliers in the finite subsets $X$ and $Y$. Notably, \textit{the Wasserstein distance is undefined if one of the finite subsets is empty}. Consequently, unlike the OSPA track-to-track distance in \eqref{def:ospa2_base}, the time-average (track-to-track) Wasserstein distance between two tracks is undefined if one track is fragmented (as shown in Fig.~\ref{fig:track_def}). One method to define a Wasserstein track-to-track distance is to embed the time information as a part of multi-object state: for any $t,u \in \mathbb{T}$, we form finite sets $\tilde{X}^{(t)} = \{ (t(k), \alpha k):  k \in \mathcal{D}^{(t)} \}$ and $\tilde{X}^{(u)} = \{ (u(k), \alpha k):  k \in \mathcal{D}^{(u)} \}$ and define
\begin{align}
    \tilde{d}_W(t,u) = d^{(p)}_W(\tilde{X}^{(t)},\tilde{X}^{(u)}),
\end{align}
where $\alpha$ is a user-defined parameter to calibrate the influence of time on the overall distance. In our experiments, we set $\alpha = 20$ m/s---the maximum velocity of all objects. Notably, for the CA-PHD-GCI fusion strategy, the underlying PHD filter does not report labels for multi-object state estimates; hence, 
tracking performance---in terms of OSPA\textsuperscript{(2)}---cannot be assessed. Thus, we only report OSPA for CA-PHD-GCI and use blue
to represent estimated object locations without labels in Fig.~\ref{fig:Scenario10_est_vs_truth_settings}b and Fig.~\ref{fig:Scenario1_est_vs_truth_settings}b.

\subsection{Scenario 1 --- two nodes with a small number of objects}

In this setting, we investigate the simple problem of tracking three mobile objects using two sensor nodes with limited FoV sensors in a  $[-500,1500]$~m $\times~[0,800]$~m area. The two nodes  are located at $[0,400]^{T}$~m and $[0,800]^{T}$~m, respectively. At each node, the sensor can only detect objects  within its relative angle of $[-50^{\circ},50^{\circ}]$ with $P_{D}=0.98$ and $r_{D}=800$~m. This scenario's duration is $80$~s, with object $1$ and object $2$ staying alive for the whole period, while object $3$ is born at time $10$~s and dies at time $60$~s. The track consensus is performed
over $5$ scans, with track length threshold $\mathcal{C}_{\text{len}}=2$, \ie,~ only unmatched tracks with lengths exceeding $2$ are retained. The considered scenario setting is shown in Fig.~\ref{fig:Scenario10_est_vs_truth_settings}a.

\begin{figure}[!tb]
\centering \includegraphics[width=0.47\textwidth]{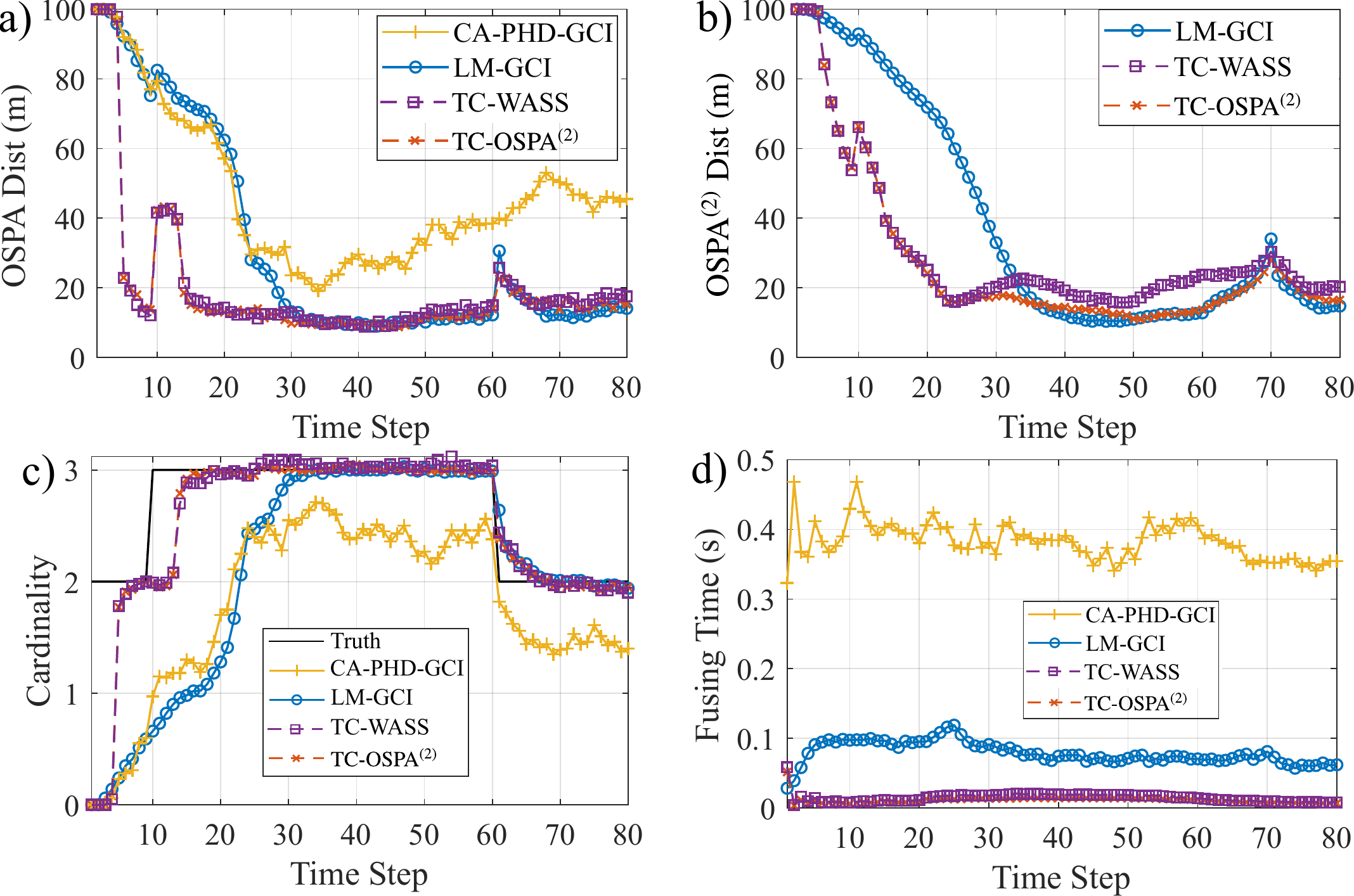}
\vspace{-0.2cm} 
\caption{Comparison results for Scenario 1: a) OSPA distance; b) OSPA\protect\textsuperscript{(2)} distance; c) Cardinality estimations; d) Fusing times.}
\vspace{-0.4cm}
\label{fig:Scenario10_results} 
\end{figure}

\begin{table}[!tb]
\centering \caption{Scenario 1 results comparison over 100 Monte Carlo runs} 
\vspace{-0.2cm}
 \label{tab:comparsion_results_s10} 
\begin{tabular}{|l|c|c|c|}
\hline 
\multicolumn{1}{|c|}{\textbf{Strategies}} & \multicolumn{1}{c|}{\textbf{OSPA (m)}} & \multicolumn{1}{c|}{\textbf{OSPA}\textsuperscript{(2)}\textbf{ (m)}} & \multicolumn{1}{c|}{\textbf{}
\begin{tabular}{@{}c@{}}
\textbf{Fusing time (s)}\tabularnewline
\end{tabular}}\tabularnewline
\hline 
CA-PHD-GCI  & 46.6  & \hl{-}  & 0.383 \tabularnewline
\hline 
LM-GCI  & 31.7  & 38.8  & 0.079 \tabularnewline
\hline 
TC-WASS  & 19.6  & 29.7  & 0.013 \tabularnewline
\hline 
TC-OSPA\textsuperscript{(2)}  & \textbf{19.0}  & \textbf{27.0}  & \textbf{0.011} \tabularnewline
\hline 
\end{tabular}
\end{table}

Fig.~\ref{fig:Scenario10_est_vs_truth_settings}, Fig.~\ref{fig:Scenario10_results} and Table~\ref{tab:comparsion_results_s10} compare results across CA-PHD-GCI, LM-GCI, TC-WASS, and TC-OSPA\textsuperscript{(2)} fusion methods. The results confirm that our proposed TC-OSPA\textsuperscript{(2)} method can accurately detect and track all three objects with consistent labels whilst consuming the shortest fusing time. Due to the small number of objects, TC-WASS also attains a low OSPA error; however, TC-WASS occasionally fails to assign correct labels and results in a higher OSPA\textsuperscript{(2)} error as seen in Table~\ref{tab:comparsion_results_s10}. In contrast, LM-GCI can only detect and track objects accurately within a sensor's own FoV. After the 30~s duration mark, all three objects are within the intersection of the two sensors' FoVs, and we can observe Node~2 to correctly track all three objects as illustrated in Fig.~\ref{fig:Scenario10_results}a and \ref{fig:Scenario10_results}c. 
Further, although CA-PHD-GCI can estimate most of these three objects, its accuracy based on the OSPA distance is the poorest because the underlying PHD filter does not perform as well as the LMB filter at each local node~\cite{hendrik2015the}. 

\subsection{Scenario 2 --- two nodes with a large number of objects}

\begin{figure}[!tb]
\centering \includegraphics[width=0.30\textwidth]{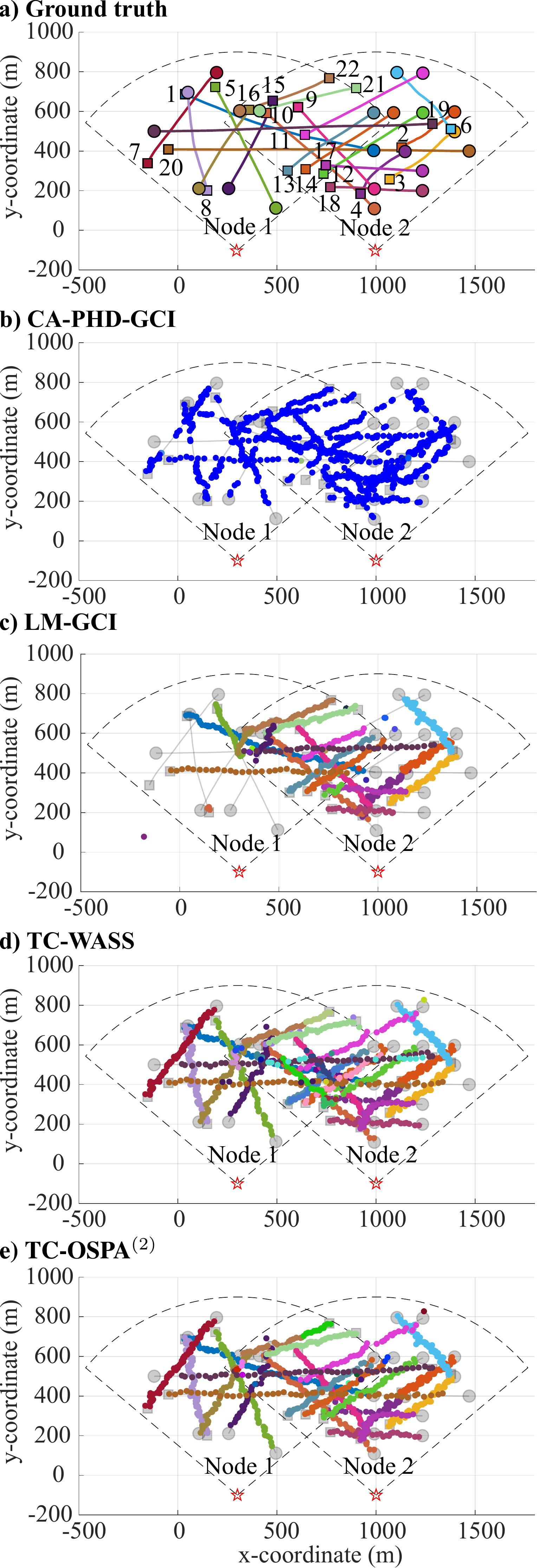}
\vspace{-0.3cm}
\caption{Scenario 2 ground-truth and consensed labelled multi-object state estimates at sensor node $2$: a) ground-truth; b) CA-PHD-GCI~(unlabelled approach); c) LM-GCI; d) TC-WASS; e) TC-OSPA\protect\textsuperscript{(2)}. Starting and stopping positions are denoted by $\circ$ and $\Box$, respectively. Colour coding represents labels of the objects.  Although less apparent, several false tracks are visible in the region of $[500,1000]$~m $\times[200,600]$~m in Fig.~\ref{fig:Scenario1_est_vs_truth_settings}d when using TC-WASS.}
\vspace{-0.7cm}
\label{fig:Scenario1_est_vs_truth_settings} 
\end{figure}

In this scenario, we consider a more challenging problem using two limited-FoV sensor nodes to track a time-varying and an unknown number of mobile objects in a  $[-500,1800]$~m $\times~[-100,1000]$~m surveillance area. The two nodes are located at $[300,-100]^{T}$~m and $[1000,-100]^{T}$~m, respectively. Each sensor can only detect objects within its limited FoV defined by a relative angle of interval $[-50^{\circ},50^{\circ}]$ with $P_{D}=0.98$ and detection range $r_{D}=1000$~m. This scenario's duration is $80$~s, with various birth and death events and a maximum of $22$ objects. The track consensus between two nodes is performed over $5$ scans, with track length threshold $\mathcal{C}_{\text{len}}=2$. The considered scenario is illustrated in Fig.~\ref{fig:Scenario1_est_vs_truth_settings}a.

\begin{figure}[!tb]
\centering \includegraphics[width=0.47\textwidth]{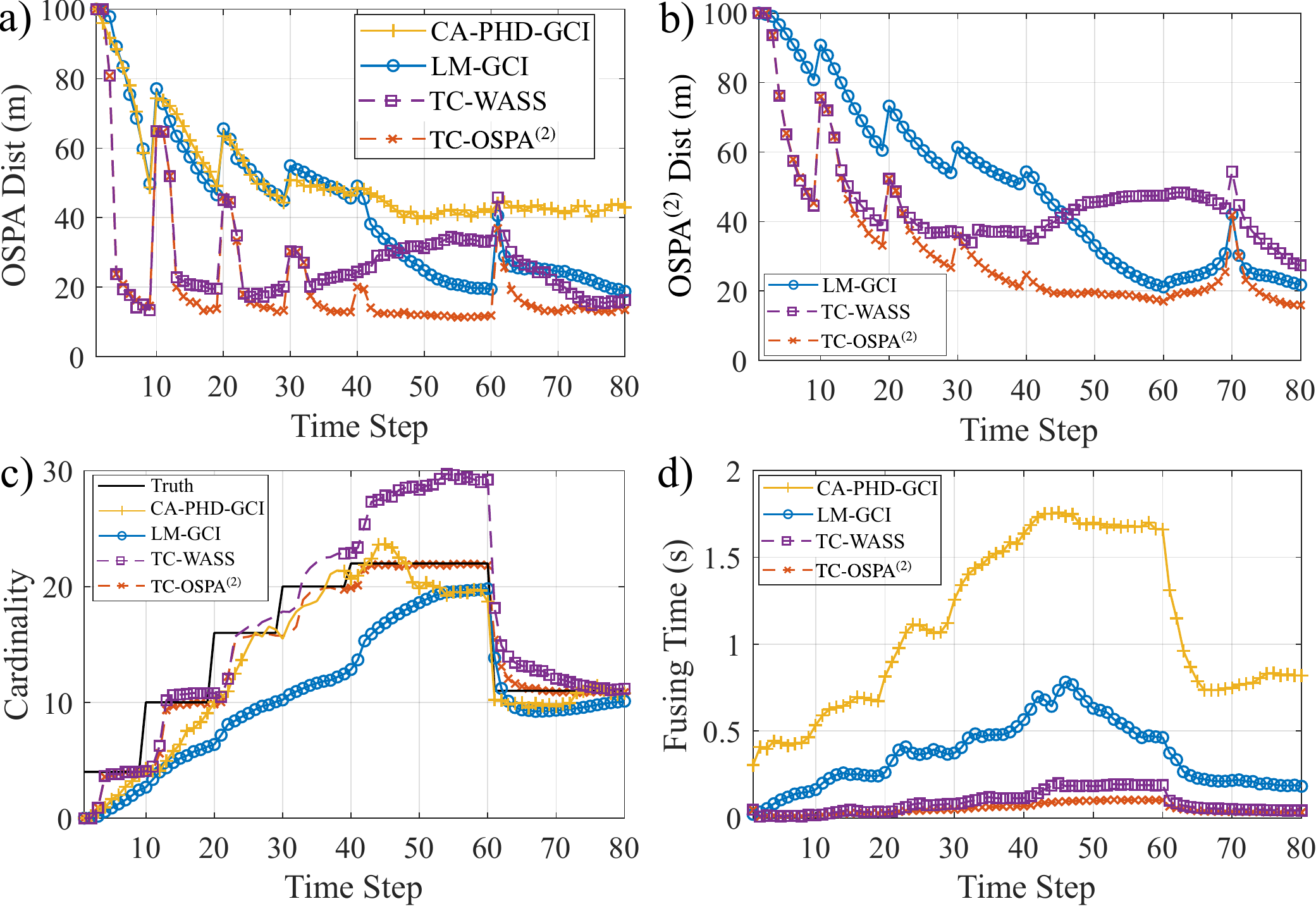}
\vspace{-0.2cm}
\caption{Comparison results for Scenario 2: a) OSPA distance; b) OSPA\protect\textsuperscript{(2)} distance; c) Cardinality estimations; d) Fusing times. }
\label{fig:Scenario1_results} 
\end{figure}

Fig.~\ref{fig:Scenario1_est_vs_truth_settings}b-d depict the consensed labelled multi-object state estimates and the true trajectories at sensor node~2 of a particular run for CA-PHD-GCI, LM-GCI, TC-WASS and TC-OSPA\textsuperscript{(2)}, respectively. The results confirm that TC-OSPA\textsuperscript{(2)} successfully detects and tracks all objects without any label inconsistency, regardless of whether objects are in the node's FoV or not. In contrast, LM-GCI can only detect and track most of the objects within the node's FoV. Although CA-PHD-GCI can detect all objects, even those outside the node's FoV, CA-PHD-GCI demonstrates the worst estimation performance as shown in Table~\ref{tab:comparsion_results_s1}. Fig~\ref{fig:Scenario1_results}c shows the cardinality estimates; the evidence therein further supports the above observation. It is expected that LM-GCI fails to detect all of the objects compared to CA-PHD-GCI, TC-WASS and TC-OSPA\textsuperscript{(2)} since only the latter three strategies are designed to cope with limited FoV sensors.

Fig.~\ref{fig:Scenario1_results}a-d and Table~\ref{tab:comparsion_results_s1} present the performance comparisons among the four fusion strategies in terms of OSPA, OSPA\textsuperscript{(2)} and fusing time over 100 Monte Carlo runs. We can see that TC-OSPA\textsuperscript{(2)} outperforms other fusion strategies, including our track consensus fusion strategy employed with the adaptation of the Wasserstein metric (TC-WASS) by large margins whilst requiring the shortest fusing time. The reason is that TC-OSPA\textsuperscript{(2)} minimises label inconsistency for limited FoV sensors while fusing the most certain object state estimates within a sensor's FoV (local labelled multi-object state estimates) to reach consensus in both position and label estimations.

\begin{table}[!tb]
\centering 
\caption{Scenario 2 comparison results over 100 Monte Carlo runs} 
\vspace{-0.2cm}
\label{tab:comparsion_results_s1} 
\begin{tabular}{|l|c|c|c|}
\hline 
\multicolumn{1}{|c|}{\textbf{Strategies}} & \multicolumn{1}{c|}{\textbf{OSPA (m)}} & \multicolumn{1}{c|}{\textbf{OSPA}\textsuperscript{(2)}\textbf{ (m)}} & \multicolumn{1}{c|}{\textbf{}
\begin{tabular}{@{}c@{}}
\textbf{Fusing time (s)}\tabularnewline
\end{tabular}}\tabularnewline
\hline 
CA-PHD-GCI  & 51.5  & \hl{-}  & 1.095 \tabularnewline
\hline 
LM-GCI  & 42.9  & 50.1  & 0.359 \tabularnewline
\hline 
TC-WASS  & 28.6  & 45.6  & 0.078 \tabularnewline
\hline 
TC-OSPA\textsuperscript{(2)}  & \textbf{21.0}  & \textbf{32.5}  & \textbf{0.052} \tabularnewline
\hline 
\end{tabular}\vspace{-0.2cm}
 
\end{table}

\begin{table}[!tb]
\centering \vspace{-0.2cm}
\caption{Performance comparison for different $P_{D}$ values obtained over 100 Monte Carlo runs for each $P_{D}$ setting and method}
\vspace{-0.2cm}
\label{tab:s1_pd} 
\begin{tabular}{|l|c|c|c|c|c|}
\cline{2-6} \cline{3-6} \cline{4-6} \cline{5-6} \cline{6-6} 
\multicolumn{1}{l|}{} & \multicolumn{1}{c|}{$P_{D}$} & \multicolumn{1}{c|}{\textbf{}
\begin{tabular}{@{}c@{}}
\textbf{CA-}\tabularnewline
\textbf{PHD-GCI}\tabularnewline
\end{tabular}} & \multicolumn{1}{c|}{\textbf{}
\begin{tabular}{@{}c@{}}
\textbf{LM-}\tabularnewline
\textbf{GCI}\tabularnewline
\end{tabular}} & \multicolumn{1}{c|}{\textbf{}
\begin{tabular}{@{}c@{}}
\textbf{TC-}\tabularnewline
\textbf{WASS}\tabularnewline
\end{tabular}} & \multicolumn{1}{c|}{\textbf{}
\begin{tabular}{@{}c@{}}
\textbf{TC-}\tabularnewline
\textbf{OSPA}\textsuperscript{(2)}\tabularnewline
\end{tabular}}\tabularnewline
\hline 
\multicolumn{1}{|l|}{\multirow{3}{*}{\textbf{OSPA (m)}}} & 0.7  & 78.9  & 61.1  & 39.9  & \textbf{37.1} \tabularnewline
\cline{2-6} \cline{3-6} \cline{4-6} \cline{5-6} \cline{6-6} 
\multicolumn{1}{|l|}{} & 0.8  & 71.1  & 52.4  & 34.5  & \textbf{30.0} \tabularnewline
\cline{2-6} \cline{3-6} \cline{4-6} \cline{5-6} \cline{6-6} 
\multicolumn{1}{|l|}{} & 0.9  & 64.4  & 44.7  & 31.1  & \textbf{24.1} \tabularnewline
\hline 
\multicolumn{1}{|l|}{\multirow{3}{*}{\textbf{OSPA}\textsuperscript{(2)}\textbf{ (m)}}} & 0.7  & \hl{-}  & 67.4  & 55.7  & \textbf{49.3} \tabularnewline
\cline{2-6} \cline{3-6} \cline{4-6} \cline{5-6} \cline{6-6} 
\multicolumn{1}{|l|}{} & 0.8  & \hl{-}  & 59.6  & 51.8  & \textbf{42.4} \tabularnewline
\cline{2-6} \cline{3-6} \cline{4-6} \cline{5-6} \cline{6-6} 
\multicolumn{1}{|l|}{} & 0.9  & \hl{-}  & 52.3  & 48.3  & \textbf{35.9} \tabularnewline
\hline 
\end{tabular}\vspace{-0.2cm}
 
\end{table}

Table~\ref{tab:s1_pd} further compares the performance of TC-OSPA\textsuperscript{(2)} with LM-GCI, CA-PHD-GCI, and TC-WASS under different detection probability, $P_{D}$, settings. It is expected that the occurrences of label inconsistency will increase when $P_{D}$ decreases. Consequently, the tracking accuracy decreases when $P_{D}$ decreases, concurring with our observation in Remark~\ref{remark:pD}. Nevertheless, the results in Table~\ref{tab:s1_pd} show the proposed TC-OSPA\textsuperscript{(2)} approach to consistently outperform the other three fusion strategies across the $P_{D}$ values.

\begin{figure}[!tb]
\centering \includegraphics[width=0.3\textwidth]{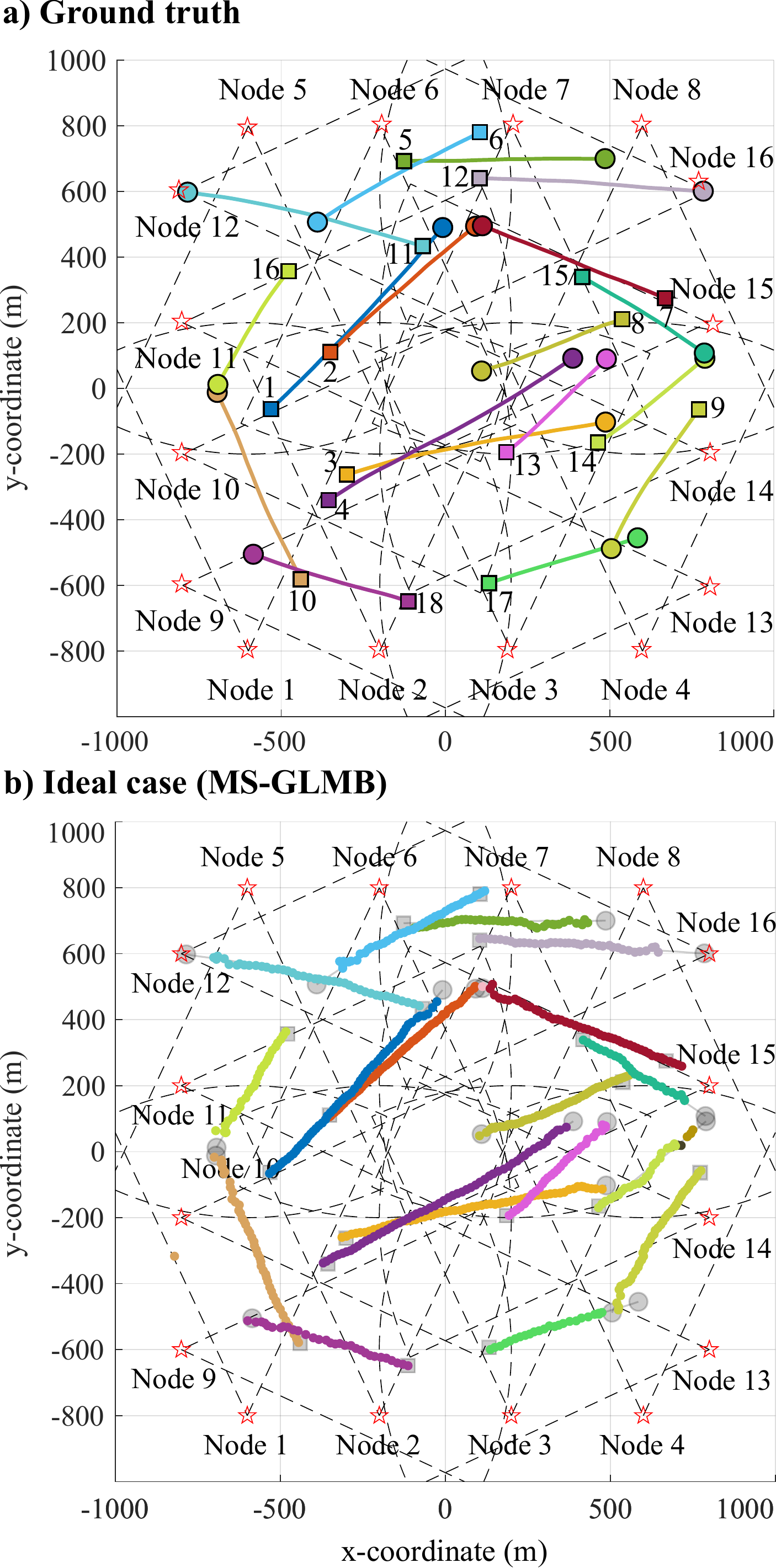}
\vspace{-0.2cm}
\caption{Scenario 3 ---  a) ground-truth; b) ideal case (MS-GLMB---centralised method where all of the sensor FoVs are completely known).  Starting and stopping positions are denoted by $\circ$ and $\Box$, respectively. Colour coding represents objects' labels.} 
\vspace{-0.6cm}
\label{fig:Scenario2_truth_ideal} 
\end{figure}
\begin{figure}[!tb]
\centering \includegraphics[width=0.3\textwidth]{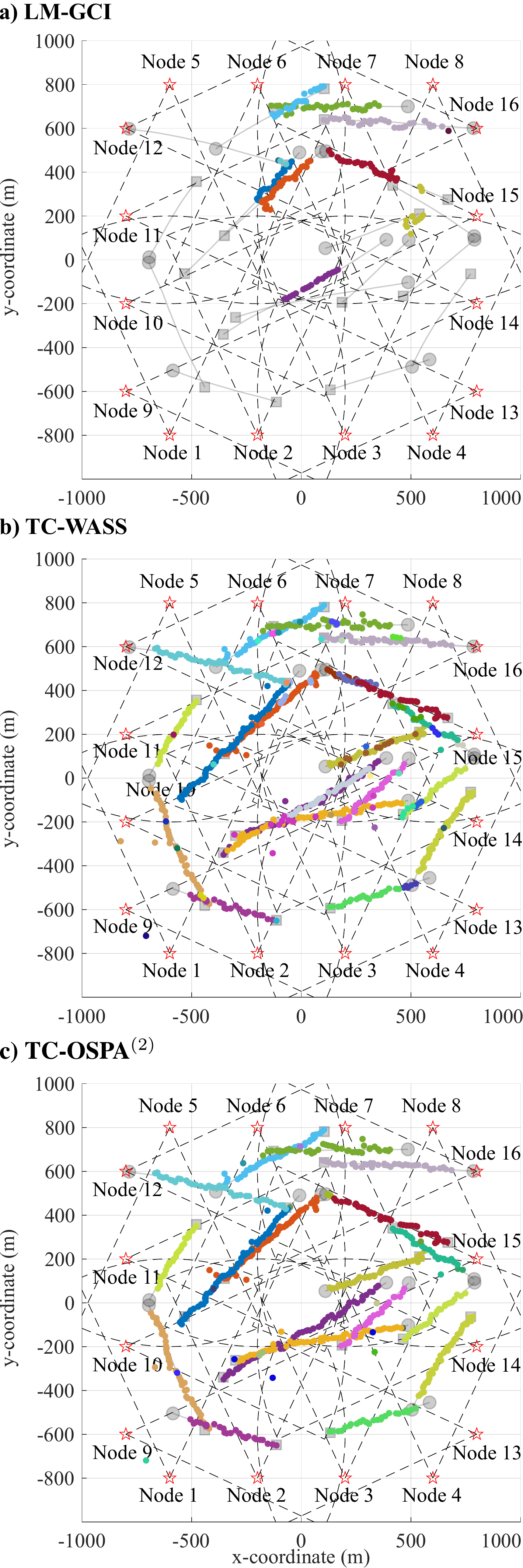}
\vspace{-0.2cm}
\caption{Scenario 3 --- consensed labelled multi-object state estimates at sensor node $7$: a) LM-GCI; b) TC-WASS; c) TC-OSPA\protect\textsuperscript{(2)}.  Starting and stopping positions are denoted by $\circ$ and $\Box$, respectively. Colour coding represents objects' labels. }
\label{fig:Scenario2_est} 
\end{figure}

\subsection{Scenario 3 --- a large number of nodes}

To further demonstrate our proposed fusion method's effectiveness, we consider a scenario with $16$ limited FoV sensor nodes for tracking a time-varying and unknown number of mobile objects in a  $[-1000,1000]$~m $\times~[-1000,1000]$~m surveillance area. These 16 nodes are positioned near the edge of the area. Each sensor can only detect objects within its limited FoV defined by a relative angle of interval $[-25^{\circ},25^{\circ}]$ with $P_{D}=0.98$ and detection range $r_{D}=1000$~m. This scenario's duration is $75$~s with various birth, death events and a maximum of $18$ objects. Further, due to higher uncertainty, the track consensus is performed over $10$ scans (instead of $5$ as in Scenario~2), with a track length threshold $\mathcal{C}_{\text{len}}=4$.  The scenario described is depicted in Fig.~\ref{fig:Scenario2_truth_ideal}a. Notably, for this scenario, we can only compare TC-OSPA\textsuperscript{(2)} with TC-WASS  and LM-GCI since it is unclear how CA-PHD-GCI can be implemented for more than two nodes. Additionally, we implement MS-GLMB in \cite{vo2019multisensor} for this scenario (see Fig.~\ref{fig:Scenario2_truth_ideal}b) as a lower bound (\textit{ideal case}) on the estimation error, \ie~ to show how close/far of our proposed fusion method is from the ideal case of centralised tracking when all of the sensors' FoVs are known.

\begin{figure}[!tb]
\centering \includegraphics[width=0.47\textwidth]{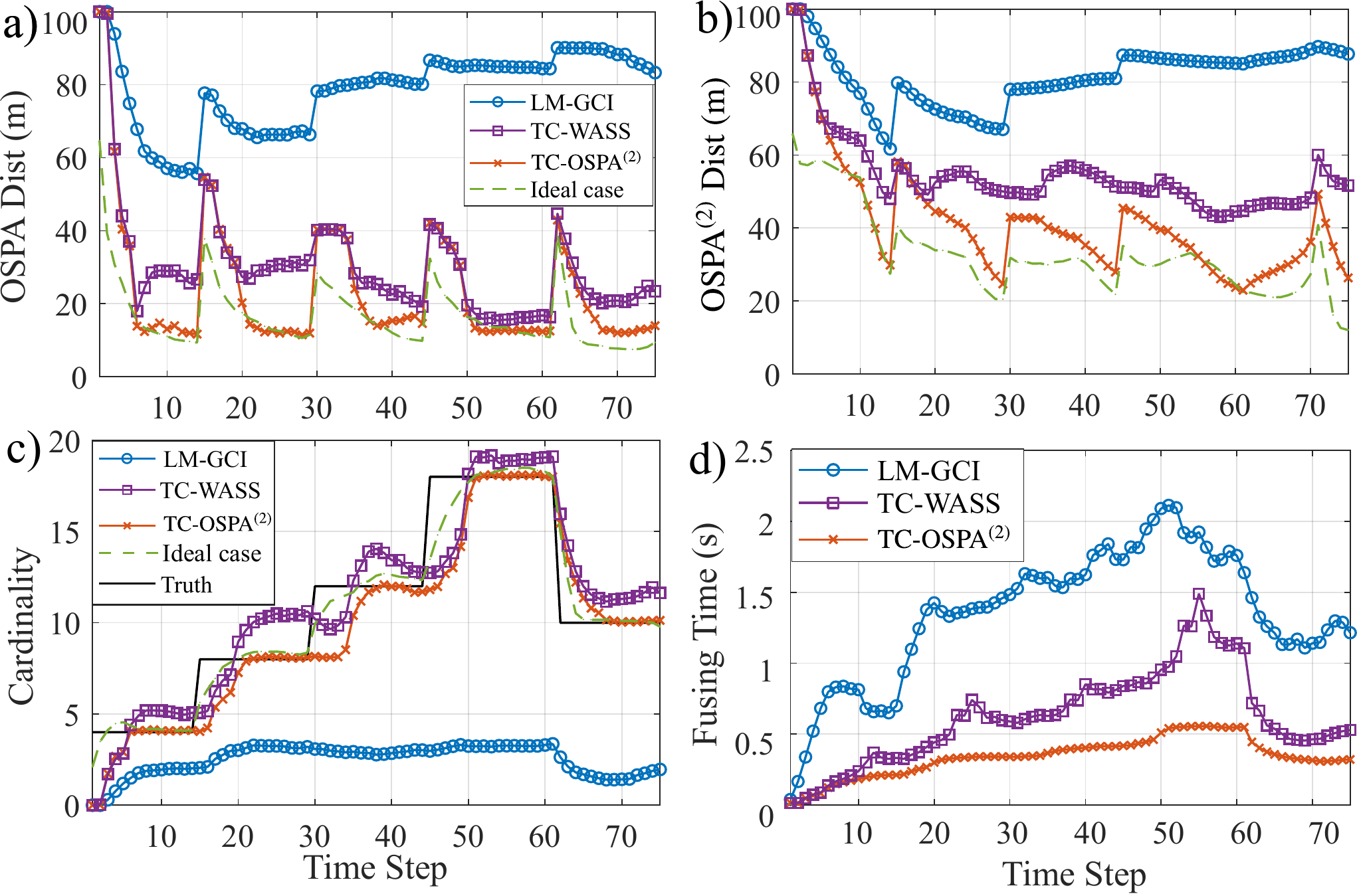}
 \caption{Comparison results at node $7$ for Scenario 3: a) OSPA distance; b) OSPA\protect\textsuperscript{(2)} distance; c) Cardinality estimations; d) Fusing times.}
\label{fig:Scenario2_results} 
\end{figure}

Fig.~\ref{fig:Scenario2_est}a-c depict the consensed labelled multi-object state estimates and the ground-truth at node $7$ for LM-GCI, TC-WASS and TC-OSPA\textsuperscript{(2)}, respectively, for one particular execution. Although LM-GCI detects a few objects outside of node $7$'s FoV, many objects are missed because LM-GCI is not designed for partially overlapped sensor FoV situations. In contrast,  TC-WASS and TC-OSPA\textsuperscript{(2)} employing our track consensus-based fusion method can detect, track and assign correct labels for most of the objects, regardless of the objects' locations. The cardinality estimates further confirm the results plotted in Fig.~\ref{fig:Scenario2_results}c, which demonstrates that TC-OSPA\textsuperscript{(2)} and TC-WASS can detect and track all  $18$ objects in this scenario. In comparison, LM-GCI can only detect up to $4$  objects on average, over 100 MC runs.

\begin{table}[!tb]
\centering \caption{Scenario 3 comparison results over 100 Monte Carlo runs} 
\label{tab:comparsion_results_s2} 
\begin{tabular}{|l|c|c|c|}
\hline 
\textbf{Strategies}  & \multicolumn{1}{l|}{\textbf{OSPA (m)}} & \multicolumn{1}{l|}{\textbf{OSPA}\textsuperscript{(2)}\textbf{ (m)}} & \multicolumn{1}{l|}{\textbf{}
\begin{tabular}{@{}l@{}}
\textbf{Fusing times (s)}\tabularnewline
\end{tabular}}\tabularnewline
\hline 
LM-GCI  & 78.7  & 81.6  & 1.35 \tabularnewline
\hline 
TC-WASS  & 30.2  & 54.4  & 0.62 \tabularnewline
\hline 
TC-OSPA\textsuperscript{(2)}  & \textbf{23.7}  & \textbf{41.3}  & \textbf{0.34} \tabularnewline
\hline 
\begin{tabular}[c]{@{}l@{}}Ideal case\\ (centralised method)\end{tabular}  & 16.6  & 32.7  & - \tabularnewline
\hline 
\end{tabular}
\end{table}

Fig.~\ref{fig:Scenario2_results}a-d provide detailed tracking performance comparisons, not easily visible in Fig.~\ref{fig:Scenario2_est}. The results further demonstrate the robustness of TC-OSPA\textsuperscript{(2)}. It significantly outperforms LM-GCI and TC-WASS across three performance metrics: OSPA, OSPA\textsuperscript{(2)} and fusing time. Table~\ref{tab:comparsion_results_s2} summarises performance comparison results confirming the effectiveness of our proposed distributed fusion strategy. It also demonstrates that the performances of our distributed fusion method closely approach the performance of the centralised method.

\begin{figure}[!tb]
\centering \includegraphics[width=0.4\textwidth]{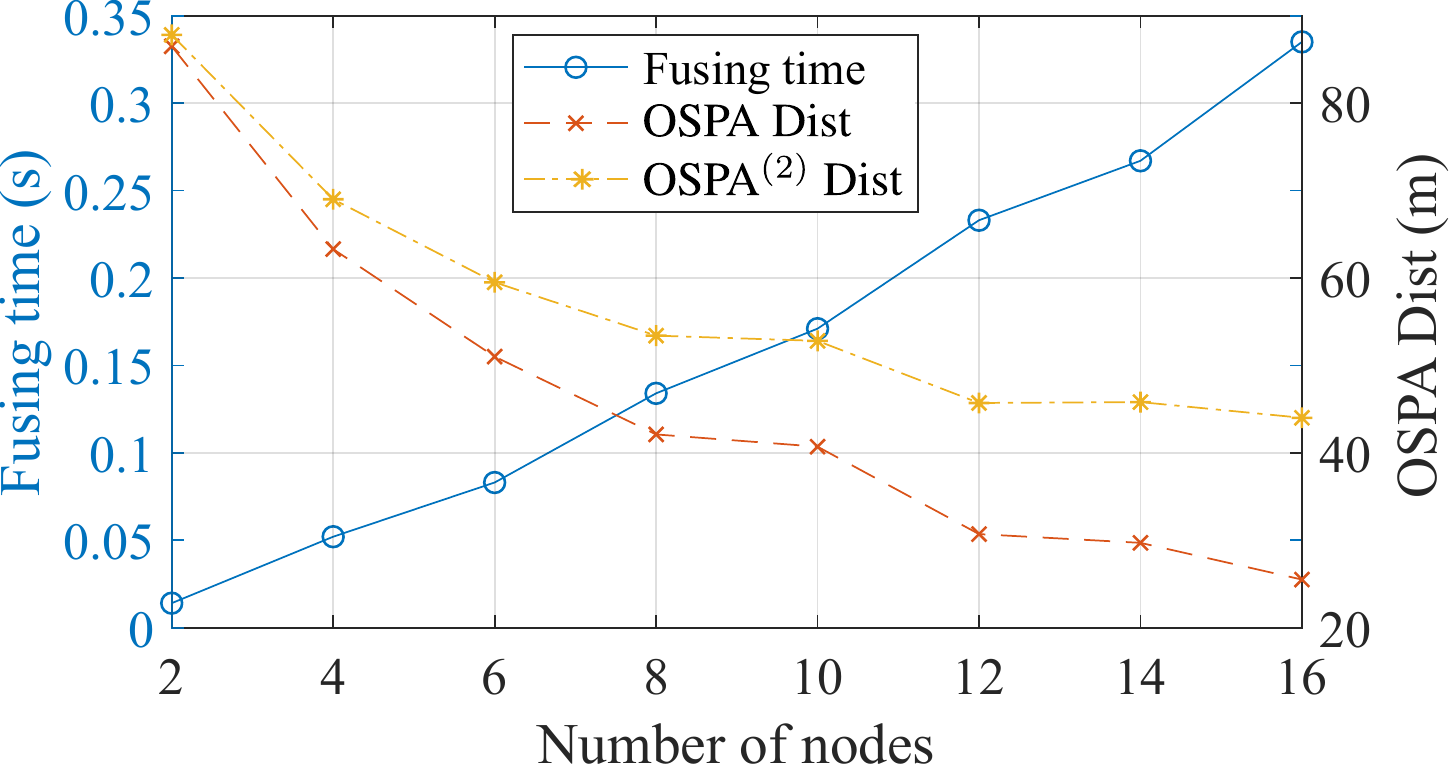}
\caption{TC-OSPA\protect\textsuperscript{(2)} tracking performance at
sensor node $2$ for Scenario 3 over 100 MC runs when the number of nodes is increased from $2$ to $16$ (\ie, the number of communications from peer nodes at node $2$ is increased from $1$ to $15$) whilst the ground truth is maintained as shown in~Fig.~\ref{fig:Scenario2_truth_ideal}a.}
 \label{fig:Scenario2_fusingtime} 
\end{figure}

Fig.~\ref{fig:Scenario2_fusingtime} plots the overall tracking performance at sensor node $2$ in Scenario 3 for TC-OSPA\textsuperscript{(2)} as the number of nodes is increased from $2$ to $16$ (\ie, the number of communications from peer nodes at node $2$ is increased from $1$ to $15$)  whilst maintaining the same ground truth as in~Fig.~\ref{fig:Scenario2_truth_ideal}a. The results validate the scalability of our proposed fusion strategy, wherein the fusing time increases linearly with respect to the number of nodes, \ie,~$\mathcal{O}(|\mathcal{N}|)$. Even with $16$ nodes, the fusing time is relatively short, suitable for real-time tracking in several applications. As expected, when the number of nodes increases, the OSPA and OSPA\textsuperscript{(2)} errors decrease since the nodes can benefit from shared local labelled multi-object state estimates of other nodes to complement their own limited FoVs, thereby improve coverage area, and tracking accuracy.

\section{Conclusion}\label{sec:conclusion} 
A new scalable DMOT solution for multi-sensors with limited FoV sensors has been proposed. Our solution consists of a novel track consensus algorithm coupled with a label consensus method based on the OSPA\textsuperscript{(2)} metric. The better efficiency and accuracy of the proposed DMOT solution are due to the fusion of multi-object state estimates using track consensus over several scans, whereas current solutions fuse multi-object densities from a single scan. Experimental results demonstrate improvements in both speed and accuracy over current methods. 

Importantly, our DMOT approach does not rely on any specific tracking methodology but only requires that each node provide a set of estimated tracks. Hence, it provides the flexibility for different network nodes to run different MOT algorithms, which is often the case in heterogeneous and ad-hoc networks.

\appendix

\subsection{Mathematical proofs}\label{sec:math_proofs}

\noindent \textbf{Proof of Proposition 1:} For any $m\in\{1,\dots,N\}$, applying triangle inequality, we have:
\begin{align}
\tilde{d}^{(c)}(a^{(m)},b^{(m)}) & \leq\tilde{d}^{(c)}(a^{(m)},u^{(m)})+\tilde{d}^{(c)}(u^{(m)},b^{(m)}),\nonumber \\
 & \leq2\mathcal{E}.
\end{align}
Similarly, using the triangle inequality and the bound $\mathcal{E}$, 
for any $n,m\in\{1,\dots,N\}$ and $n\neq m$,
we have 
\begin{align}
\tilde{d}^{(c)}(a^{(m)},b^{(n)}) & \geq\tilde{d}^{(c)}(u^{(m)},b^{(n)})-\tilde{d}^{(c)}(u^{(m)},a^{(m)})\\
 & \geq\tilde{d}^{(c)}(u^{(m)},b^{(n)})-\mathcal{E},\nonumber \\
 & \geq\tilde{d}^{(c)}(u^{(m)},u^{(n)})-2\mathcal{E}.
\end{align}
Using \eqref{eq:label_consistent_condi} and $\tilde{d}^{(c)}(u^{(m)},u^{(n)})>4\mathcal{E}$, we have 
\begin{align}
\tilde{d}^{(c)}(a^{(m)},b^{(n)}) & >\tilde{d}^{(c)}(a^{(m)},b^{(m)}).\label{eq:mn_neq}
\end{align}
Using Definition~\ref{def:solution_of_optimal_matching} and \eqref{eq:remaing_tracks_dist}, we have 
\begin{align}
&\pi^{*}  =\argmin_{\pi\in\Pi_{|\mathbf{T}^{(b)}_k|} } \sum_{m=1}^{|\mathbf{T}_{k}^{(a)}|}\tilde{d}^{(c)}(a^{(m)},b^{(\pi(m))})\\ 
 & =\argmin_{\pi\in\Pi_{|\mathbf{T}^{(b)}_k|} }\Big[\sum_{m=1}^{N}\tilde{d}^{(c)}(a^{(m)},b^{(\pi(m))})+\big(|\mathbf{T}_{k}^{(a)}|-N\big)c\Big].\label{eq:optimal_matching_new}
\end{align}
Since $\tilde{d}^{(c)}(a^{(m)},b^{(\pi(m))}) =c ~\forall \pi(m) >N$ according to \eqref{eq:remaing_tracks_dist} and $\big(|\mathbf{T}_{k}^{(a)}|-N\big)c$ is a constant,   \eqref{eq:optimal_matching_new} is equivalent to 
\begin{align}
\pi^{*} & =\argmin_{\pi \in \Pi_{N}}J(\pi),\label{eq:optimal_matching_new_2}
\end{align}
where $J(\pi)=\sum_{m=1}^{N}\tilde{d}^{(c)}(a^{(m)},b^{\pi(m)})$.

Let $\hat{\pi}\in \Pi_{N}$ be an assignment function that satisfies \eqref{eq:H_star_mm} (\ie, $\hat{\pi}(m)=m~\forall m\in\{1,\dots,N\})$. According to Definition~\ref{def:label_consistency}, if $\pi^{*}$ satisfies~\eqref{eq:H_star_mm}, and we assign the same label to each matched pair, then the $N$ distinct objects have label consistency. 

We now prove that $\pi^{*}=\hat{\pi}$ by using contradiction. Suppose that $\pi^{*}\neq\hat{\pi}$. Then according to \eqref{eq:optimal_matching_new_2}, $J(\pi^{*})<J(\hat{\pi})$. Thus, $\exists m,n\in\{1,\dots,N\}$ and $m\neq n$ such that $\pi^{*}(m)=n$ and $\tilde{d}^{(c)}(a^{(m)},b^{(n)})\leq\tilde{d}^{(c)}(a^{(m)},b^{(m)})$, otherwise we would have $J(\pi^{*})>J(\hat{\pi})$. However, $\tilde{d}^{(c)}(a^{(m)},b^{(n)})\leq\tilde{d}^{(c)}(a^{(m)},b^{(m)})$ contradicts \eqref{eq:mn_neq}. Therefore $\pi^{*}=\hat{\pi}.\quad\blacksquare$

\vspace{2mm}
\noindent \textbf{Proof of Proposition 2:} From the Definition~\ref{def:pos_px}, we have: 
\begin{align}
P_{X}(a^{(m)})=\dfrac{|\mathcal{D}^{(a^{(m)})}|}{n}=P_{X}^{\min}+\epsilon,
\end{align}
where $n=k-j+1$ and $\epsilon\geq0$.

Using \eqref{def:ospa2_base}
\begin{equation}
	\resizebox{0.5\textwidth}{!}{ $\begin{aligned}
			&\tilde{d}^{(c)}(a^{(m)},u^{(m)}) 
	= \sum_{i=j}^{k} \dfrac{d^{(c)}_{\mathtt{O}}(\{a^{(m)}(i)\},\{u^{(m)}(i)\})}{n}\\
	&=\dfrac{1}{n}\sum_{i\in \mathcal{D}^{(a^{(m)})}}d^{(c)}_{\mathtt{O}}(\{a^{(m)}(i) \},\{u^{(m)}(i)\}) + \dfrac{n-|\mathcal{D}^{(a^{(m)})}|}{n}c \\
	&\leq (P^{\min}_{X} + \epsilon)\varepsilon  + (1-P^{\min}_{X} - \epsilon)c \\
	&\leq \varepsilon  P^{\min}_{X} + (1-P^{\min}_{X})c -\epsilon(c-\varepsilon ) \\
	&\leq \varepsilon  P^{\min}_{X} + (1-P^{\min}_{X})c.
		\end{aligned}$ }
\end{equation}
Similarly, we have 
\begin{align*}
\tilde{d}^{(c)}(b^{(m)},u^{(m)})\leq\varepsilon P_{X}^{\min}+(1-P_{X}^{\min})c.\quad\blacksquare
\end{align*}

\subsection{Pseudocodes}~\label{sec:pseudo_codes}
The pseudocode for determining matched pairs based on OSPA\textsuperscript{(2)} distance between two tracks is given in Algorithm~\ref{Algo_DetermineMatchedPairs}. The pseudocode for updating matched history for the matched pairs is given in Algorithm ~\ref{Algo_UpdateMatchedHistory}. Lastly, the pseudocode for updating consensed labels to ensure label consistency is given in Algorithm~\ref{Algo_UpdateLabels}.

\subsubsection{Determine Matched Pairs --- Algorithm~\ref{Algo_DetermineMatchedPairs}}
The associated cost $C$ between two sets of tracks is initialised in line $1$ based on the size of $\mathbf{T}_k^{(a)}$ and $\mathbf{T}_k^{(b)}$, and thereafter computed in lines $2-6$ via the OSPA track-to-track distance in \eqref{def:ospa2_base}. The optimal assignment $\pi^*$ is computed in line $7$ via the Hungarian algorithm. The assignment matrix $H^*$ is initialised based on the size of $\mathbf{T}_k^{(a)}$ and $\mathbf{T}_k^{(b)}$ in line $1$ and thereafter computed in lines $8-11$ such that only optimal assignments with associated cost less than $c$ (see line $11$) are considered as optimal matches. Here, $\odot$ denotes the element-wise product between two matrices. The matched pair matrix $Q$  is computed in lines $12-15$ based on the assignment matrix $H^*$.
 
\setcounter{algorithm}{0}
\renewcommand\thealgorithm{B.\arabic{algorithm}} 
\begin{algorithm}[!htbp]
	\footnotesize
	\caption{\textsf{DetermineMatchedPairs}}\label{Algo_DetermineMatchedPairs} 
	\begin{algorithmic}[1] 
		\Statex \textbf{Input}: $~\mathbf{T}_k^{(a)};\mathbf{T}_k^{(b)};$
		\Statex \textbf{Output}: $~Q_k=[Q^{(a)}_k,Q^{(b)}_k];$
		\State $C:=\textsf{{\footnotesize {zeros(}}}|\mathbf{T}^{(a)}_{k}| , |\mathbf{T}^{(b)}_{k}|);H^* :=  \textsf{{\footnotesize {zeros(}}}|\mathbf{T}^{(a)}_{k}|, |\mathbf{T}^{(b)}_{k}|);$ 
		\For{$m=1:|\mathbf{T}^{(a)}_k|$}
		\For{$n=1:|\mathbf{T}^{(b)}_k|$}
		\State $C_{m,n} := \tilde{d}^{(c)}(a^{(m)},{b}^{(n)})$\textsf{\footnotesize { via }} \eqref{def:ospa2_base}
		\EndFor
		\EndFor
		\State $\pi^*:= $ \textsf{\footnotesize {OptimalAssignment}}$(C);$ \Comment{Use Hungarian's algorithm.}
		\For{$m=1:|\mathbf{T}^{(a)}_k|$}
		    \State $H^*_{m,\pi^*(m)} :=1;$
		\EndFor
		\State $H^*:= H^* \odot (C < c);$ \Comment{Select assignments with cost $ C < c$.}
		\State $i^{(a)}_{k} := [1 : |\mathbf{T}^{(a)}_{k}|]^T;$ $i^{(b)}_{k} := [1 : |\mathbf{T}^{(b)}_{k}|]^T;$
		\State $Q^{(a)}_k := i^{(a)}_{k}; $ $Q^{(b)}_k := H^* \cdot i^{(b)}_{k}; $ $Q_k:=[Q^{(a)}_k,Q^{(b)}_k];$
		\State $Q^{\text{check}} :=[Q^{(a)}_k \odot Q^{(b)}_k] > 0;$ \Comment{Ensure $H^*_{m,\pi^*(m)}=1$.}
		\State $Q_k:=Q_k(:,Q^{\text{check}});$
	\end{algorithmic}
\end{algorithm}

\subsubsection{Update Matched History --- Algorithm~\ref{Algo_UpdateMatchedHistory}} The matched history matrix $\Xi^{(a,b)}_{1:k}$ at time $k$ is initialised in lines $1-2$ based on the current label spaces of two nodes, and the previous matched history matrix $\Xi^{(a,b)}_{1:k-1}$ at time $k-1$. Line $3$ initialises the indices of the label spaces from two node $a$ and $b$. The matched history matrix $\Xi^{(a,b)}_{1:k}$ is computed in lines $4-10$ based on the matched pair matrix $Q_k=[Q^{(a)}_k,Q^{(b)}_k]$ from Algorithm~\ref{Algo_DetermineMatchedPairs} such that if track $i^{(m)}\in\{1,\dots,|\mathbf{L}_{1:k}^{(a)}|\}$ from node $a$ (line 7) is matched to track $i^{(n)}\in\{1,\dots,|\mathbf{L}_{1:k}^{(b)}|\}$ from node $b$ (line 8), then we add $1$ into  $\Xi^{(a,b)}_{1:k-1}(i^{(m)},i^{(n)})$ (line 9) to  record the number of instances that tracks from node $a$ is matched with track from node $b$.

\begin{algorithm}[!htbp]
	\footnotesize
	\caption{\textsf{UpdateMatchedHistory}}\label{Algo_UpdateMatchedHistory} 
	\begin{algorithmic}[1] 
		\Statex \textbf{Input}: $~\Xi^{(a,b)}_{1:k-1};~\mathbf{L}^{(a)}_{1:k};~\mathbf{L}^{(b)}_{1:k};~Q_k = [Q_k^{(a)},Q_k^{(b)}];$
		\Statex \textbf{Output}: $~\Xi^{(a,b)}_{1:k};$
		\State $\Xi^{(a,b)}_{1:k} := \textsf{\footnotesize{zeros(}}|\mathbf{L}^{(a)}_{1:k}| , |\mathbf{L}^{(b)}_{1:k}|);$
		\State $\Xi^{(a,b)}_{1:k}(1:|\mathbf{L}^{(a)}_{1:k-1}|,1:|\mathbf{L}^{(b)}_{1:k-1}|) := \Xi^{(a,b)}_{1:k-1};$
		\State $i^{(a)}_{1:k} := 1 : |\mathbf{L}^{(a)}_{1:k}|;~i^{(b)}_{1:k} := 1 : |\mathbf{L}^{(b)}_{1:k}|$
		\For{$i=1:|Q_k^{(a)}|$}
		\State $m:=Q_k^{(a)}(i);~n:=Q_k^{(b)}(i);$
		\State $\boldsymbol{\ell}^{(m)}:=\mathbf{L}^{(a)}_{k}(:m);~\boldsymbol{\ell}'^{(n)}:=\mathbf{L}^{(b)}_{k}(:,n);$
		\State $i^{(m)}:=i^{(a)}_{1:k}(\boldsymbol{\ell}^{(m)} = \mathbf{L}^{(a)}_{1:k});$ 
		\State $i^{(n)}:=i^{(b)}_{1:k}(\boldsymbol{\ell}'^{(n)} = \mathbf{L}^{(b)}_{1:k});$
		\State $\Xi^{(a,b)}_{1:k}(i^{(m)},i^{(n)}):=\Xi^{(a,b)}_{1:k}(i^{(m)},i^{(n)})+1;$
		\EndFor
	\end{algorithmic}
\end{algorithm}


\subsubsection{Update Labels--- Algorithm~\ref{Algo_UpdateLabels}} The graph $G=(V,E)$, which represents the matched history matrices amongst all nodes $(\{\Xi^{(a,b)}_{1:k}\}_{a,b \in \mathcal{N}})$, is created in line $1$. Here, $V= 1: |\{\mathbf{L}^{(a)}_{1:k}\}_{a \in \mathcal{N} }|$ is the set of vertices of the graph, and $E$ is the set of edges representing matches between labels such that if $(i,j) \in E$ then the track $\{\mathbf{L}^{(a)}_{1:k}\}_{a \in \mathcal{N} }(:,i)$ has at least one instance from time $1$ to $k$ (based on $\{\Xi^{(a,b)}_{1:k}\}_{a,b \in \mathcal{N}}$) that  is matched to track $\{\mathbf{L}^{(a)}_{1:k}\}_{a \in \mathcal{N} }(:,j)$.  Each label $\boldsymbol{\ell} \in \mathbf{L}^{\text{con}}_k $ is updated in the main loop from lines $2-18$ as follows. The vertex $m \in V $ is computed based on label $\boldsymbol{\ell}$ in line $4$, while the set of all nearest vertices $M$ connected to $m$ is computed in line $5$ using the \textsf{nearest} function in MATLAB. From the set $M$ of all connected vertices to the vertex $m$, we conduct a search using the inner loop (lines $7-17$) to find the least label from $M$ using lexicographical order defined in \eqref{eq:label_comparison} (line 10) subject to the label's uniqueness constraint in lines $12-15$.

\begin{algorithm}[!htbp]
	\footnotesize
	\caption{\textsf{UpdateLabels}}\label{Algo_UpdateLabels} 
	\begin{algorithmic}[1] 
		\Statex \textbf{Input}: $~\mathbf{L}^{\text{con}}_k;~~\{\Xi^{(a,b)}_{1:k}\}_{a,b \in \mathcal{N}};~\{\mathbf{L}^{(a)}_{1:k}\}_{a \in \mathcal{N} };$
		\Statex \textbf{Output}: $~~~\mathbf{L}^{\text{con}}_k;$
		\Statex{$\triangleright$ Create a graph representation using matched history matrices (see label  consensus in Section~\ref{sec:achieving_label_consensus}).}
		\State $G = (V,E):= $ \textsf{\footnotesize {CreateAGraph}}$(\{\Xi^{(a,b)}_{1:k}\}_{a,b \in \mathcal{N}});$ 
		\For{$i=1:|\mathbf{L}^{\text{con}}_k |$}
		\State $\boldsymbol{\ell}:=\mathbf{L}^{\text{con}}_k (:,i);$
		\State $m:=V(\boldsymbol{\ell} = \{\mathbf{L}^{(a)}_{1:k}\}_{a \in \mathcal{N} });$
		\State $M:= $ \textsf{\footnotesize {nearest}}$(G,m);$ \Comment{Get all connected vertices to vertex $m$.}
		\State $count:=|M|;$
		\While{$count > 0$}
		\State $count~:=count-1;$
		\State $\mathbf{L}^{\text{sel}}:= \{\mathbf{L}^{(a)}_{1:k}\}_{a \in \mathcal{N} }(:,M);$
		\Statex{\quad\quad\quad$\triangleright$ Get the smallest index using lexicographical order via \eqref{eq:label_comparison}.}
		\State $[\sim,n]:=\textsf{{\footnotesize {min}}}(\mathbf{L}^{\text{sel}});$ 
		\State $\boldsymbol{\ell}'=\mathbf{L}^{\text{sel}}(:,n);$
		\If{$\boldsymbol{\ell}' \notin  \mathbf{L}^{\text{con}}_k $} \Comment{Ensure labels' uniqueness.}
		\State $\mathbf{L}^{\text{con}}_k (:,i) :=\boldsymbol{\ell}';$ \Comment{Update the label.}
		\State \textsf{break;} \Comment{Escape while loop.}
		\EndIf
		\State $M(n) := [];$ \Comment{Remove $n$ from the nearest vertices.}
		\EndWhile
		\EndFor
	\end{algorithmic}
\end{algorithm}


\addcontentsline{toc}{section}{References}

%

\begin{IEEEbiography}[{\includegraphics[width=1in,height=1.25in,clip,keepaspectratio]{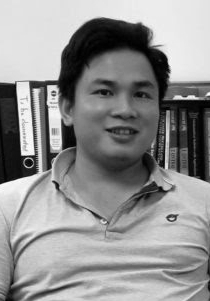}}]{Hoa Van Nguyen} received his Bachelor degree in electrical engineering from Portland State University, Oregon, U.S.A in 2012, and the PhD degree in computer science from The University of Adelaide, in 2020. He is currently a Post-Doctoral Research Fellow with the School of Computer Science, The University of Adelaide. His research interests include signal processing, robotics, multi-object tracking, and multi-sensor control.
\end{IEEEbiography}

\vspace{-0.82cm}

\begin{IEEEbiography}[{\includegraphics[width=1in,height=1.25in,clip,keepaspectratio]{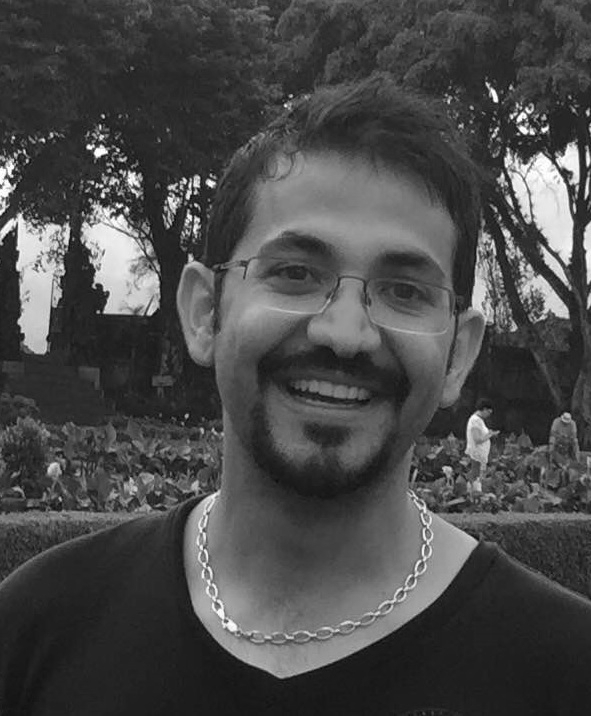}}]{Hamid Rezatofighi} is a lecturer at the Faculty of Information Technology, Monash University, Australia. Before that, he was an Endeavour Research Fellow at the Stanford Vision Lab (SVL), Stanford University and a Senior Research Fellow at the Australian Institute for Machine Learning (AIML), The University of Adelaide. His main research interest focuses on computer vision and vision-based perception for robotics, including object detection, multi-object tracking, human trajectory and pose forecasting and human collective activity recognition. He has also research expertise in Bayesian filtering, estimation and learning using point process and finite set statistics.
\end{IEEEbiography}

\vspace{-0.82cm}

\begin{IEEEbiography}[{\includegraphics[width=1in,height=1.25in,clip,keepaspectratio]{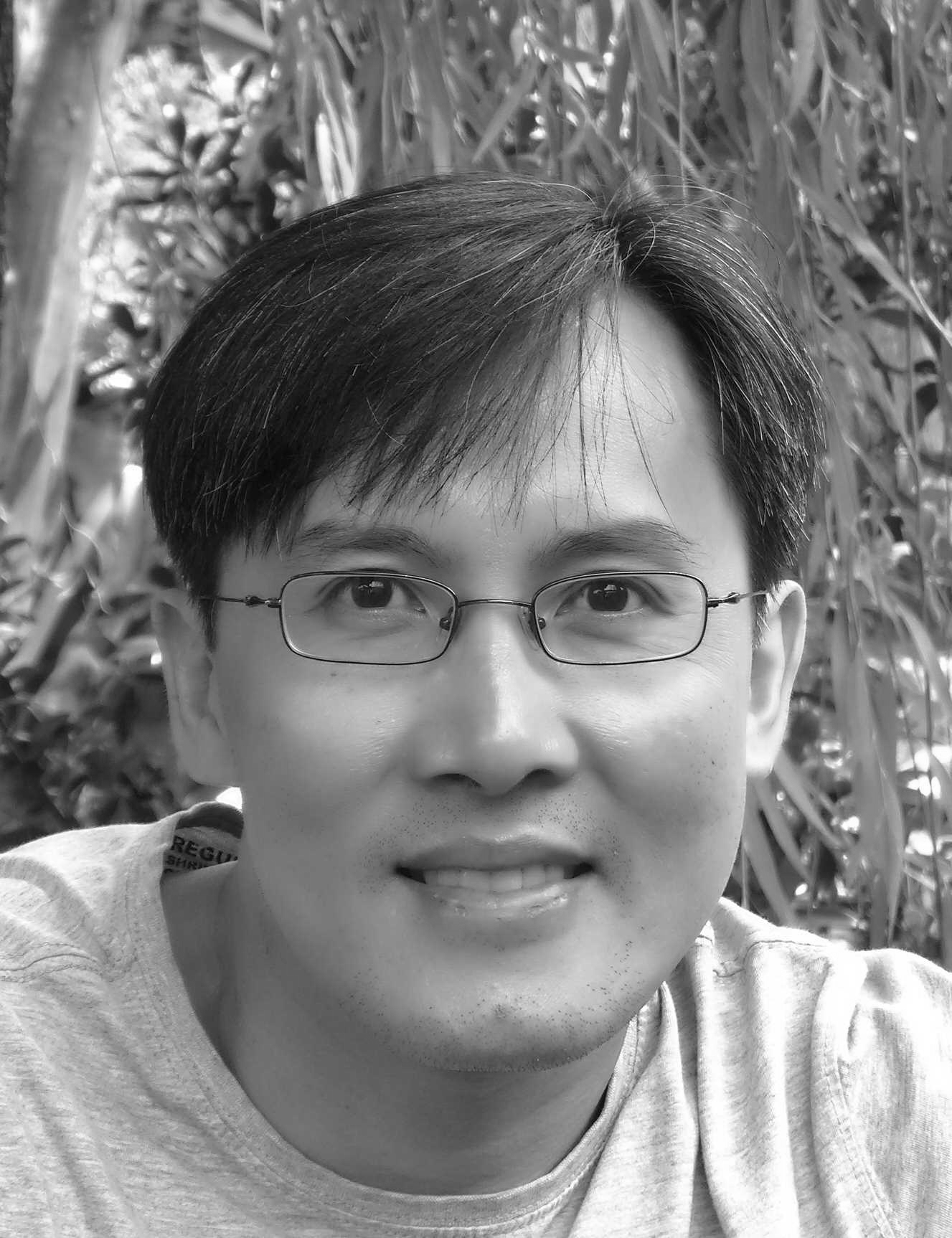}}]{Ba-Ngu Vo} received his Bachelor degrees in Mathematics and Electrical Engineering with first class honors in 1994, and PhD in 1997. Currently he is Professor of Signals and Systems at Curtin University. Vo is a Fellow of the IEEE, and is best known as a pioneer in the stochastic geometric approach to multi-object system. His research interests are signal processing, systems theory and stochastic geometry.
\end{IEEEbiography}

\vspace{-0.82cm}

\begin{IEEEbiography}[{\includegraphics[width=1in,height=1.25in,clip,keepaspectratio]{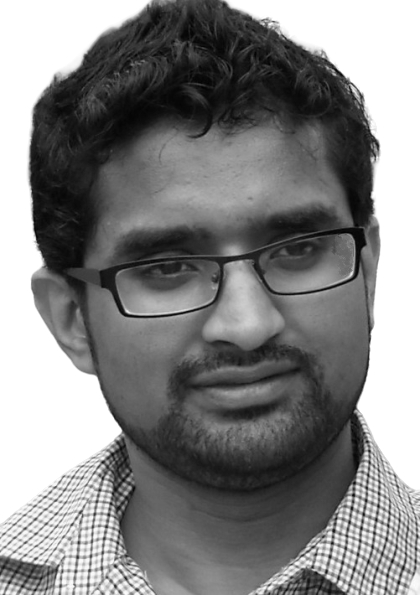}}]{Damith C. Ranasinghe}  received the Ph.D. degree in Electrical and Electronic Engineering from The University of Adelaide, Australia, in December 2007. In the past, he was a Visiting Scholar with the Massachusetts Institute of Technology, Cambridge, MA, USA, and a Post-Doctoral Research Fellow with the University of Cambridge, Cambridge, U.K. Currently, he is an Associate Professor at The University of Adelaide. His research interests lie broadly in the areas of autonomous systems, machine learning, and security.
\end{IEEEbiography}




\end{document}